\documentclass{aa}  

\usepackage{graphicx}
\usepackage{txfonts}
\begin{document}

   \title{ Ultraviolet to infrared emission of  $z>1$ galaxies:  Can we derive reliable star formation rates and stellar masses?}


   \author{V. Buat 
          \inst{1}
       \and   S. Heinis \inst{1,2}
    \and      M. Boquien \inst{1}
    \and      D. Burgarella \inst{1}
       \and V. Charmandaris\inst{3,4,5}
       \and S. Boissier \inst{1}
   \and        A. Boselli  \inst{1}
   \and D. Le Borgne \inst{6}
   \and G. Morrison \inst{7,8}}

   \institute{Aix-Marseille  Universit\'e,  CNRS, LAM (Laboratoire d'Astrophysique de Marseille) UMR7326,  13388, Marseille, France\\
              \email{veronique.buat@oamp.fr}
           \and Department of Astronomy 
CSS Bldg., Rm. 1204, Stadium Dr. 
University of Maryland 
College Park, MD 20742-2421 
\and University of Crete, Department of Physics and Institute of Theoretical \& Computational Physics, GR-71003 Heraklion, Greece
      \and IESL/Foundation for Research \& Technology-Hellas, GR-71110 Heraklion, Greece
      \and Chercheur Associ\'e, Observatoire de Paris, LERMA (CNRS:UMR8112), 61 Av. de l'Observatoire,
F-75014, Paris, France
\and  Institut d'Astrophysique de Paris, UMR 7095, CNRS, UPMC Univ. Paris 06, 98bis Boulevard Arago, F-75014 Paris, France
\and Institute for Astronomy, University of Hawaii, Honolulu, HI, 96822, USA
         \and Canada-France-Hawaii telescope, Kamuela, HI, 96743, USA
    }

   \date{Received ; accepted }

 
  \abstract
  {}
  {Our knowledge of the cosmic mass assembly relies on measurements of star formation rates (SFRs) and stellar masses ($M_{\rm star}$), of galaxies as a function of redshift. These parameters must be estimated in a consistent way with a good knowledge of systematics before studying their correlation and the variation of  the specific star formation rate. Constraining these fundamental properties of galaxies across the Universe is of utmost importance if we want to understand galaxy formation and evolution.}
   {We seek to derive star formation rates and stellar masses in distant galaxies and  to  quantify the main uncertainties  affecting their measurement. We explore the impact of  the assumptions  made  in  their derivation with standard calibrations or through a fitting process, as well as   the impact of  the   available data, focusing on the role of infrared (IR) emission originating from dust.}
   {We build  a sample of galaxies with $z>1$, all   observed from the ultraviolet   to the infrared  in their rest frame. The data are fitted with the code CIGALE, which is also used to   build and analyse a  catalogue of mock galaxies.   Models with different star formation histories are introduced:  an exponentially decreasing or  increasing star formation rate and  a more complex one coupling a decreasing star formation rate   with a younger burst of   constant star formation.  We define different set of data, with or without a good sampling of the ultraviolet range, near-infrared, and thermal  infrared  data. Variations of the  metallicity are also investigated. The impact of these different cases on the determination of stellar mass and  star formation rate  are analysed.}
   {Exponentially decreasing models with a redshift formation of  the stellar population $z_{\rm f} \simeq 8$  cannot fit  the data correctly. All the other models  fit  the data  correctly at the price of unrealistically young ages  when the age of the single stellar population is taken to be a  free parameter, especially for the exponentially decreasing models. The best fits are obtained with  two stellar populations. As long as one measurement of the  dust emission continuum  is available, SFR  are robustly estimated whatever the chosen model is,  including standard recipes. The  stellar mass  measurement is more subject to uncertainty, depending on the chosen model and the presence of near-infrared  data, with an  impact on the SFR-$M_{\rm star}$ scatter plot.  Conversely,  when thermal infrared  data from dust emission are missing, the uncertainty on SFR measurements largely exceeds that of stellar mass. Among all physical properties investigated here, the stellar ages are found to be  the most difficult to constrain and this uncertainty acts as a second  parameter in SFR measurements and as  the most important parameter for stellar mass measurements. }
   {}
   
   \titlerunning{SFR and $M_{\rm star}$ determinations at z>1}
   \maketitle
%

\section{Introduction}

Star formation rates (SFR) and stellar masses ($M_{\rm star}$)  are the main parameters estimated from large samples of galaxies and they can be used to constrain their star formation history and the evolution of  their baryonic content.  A large number of works found a tight relation between SFR and $M_{\rm star}$ both at low and high redshift (e.g., \cite{brinchmann04,daddi07,elbaz07,rodighiero11}), which is often called main sequence (MS) of galaxies (\cite{noeske07}). The slope and the scatter of this relation  as well as its evolution with redshift put constraints on the star formation history of the galaxies  as a function of their mass. The galaxies  located on this  MS   may experience  a rather smooth star formation evolution during several Gyr (\cite{heinis13b}) and the starburst mode seems to  play a minor role in the  production of stars (\cite{rodighiero11}). \\

The degree to which we can interpret these observations depends on our ability to estimate SFR and $M_{\rm star}$. Two major methods (not independent) are commonly used to measure SFR and $M_{\rm star}$. The first approach consists of using empirical recipes. The SFR is  deduced by applying conversion factors between   an observed emission   coming mostly from   young stars  and   the SFR (e.g.,\cite{kennicutt98}). The impact of dust attenuation has long been identified as a major issue.  To overcome it,  some  calibrations combine different wavelengths and  account for all the star formation directly observed in ultraviolet (UV)-optical or reprocessed in  thermal infrared (IR)  (e.g.,\cite{hao11, KennEvans12,calzetti12}). These relations  rely on strong assumptions on  star formation history (\cite{boissier12,calzetti12}), which are valid for local, normal galaxies, but may well break down for more extreme objects and at high redshift (\cite{kobayashi12,schaerer13}).  
The $M_{\rm star}$ estimations are also based on tabulated relations between  mass to light ($M/L$) ratios  and colors (\cite{bell03,zibetti09}). The accuracy of their determination when rest-frame near-infrared (NIR) data are either included or not  included  remains an open issue:  optical colors-M/L relations are less sensitive to the uncertain  thermally pulsing asymptotic giant branch  evolutionary phase, but  the uncertainty due to dust reddening is strongly minimized in NIR (\cite  [and references therein]{conroy13}. The determination of SFR and $M_{\rm star}$ is  also strongly  dependent  on the choice of the stellar libraries and initial mass function (\cite{bell03,muzzin09,marchesini09}).\\
Another widespread method  to derive these physical parameters  is to exploit the full panchromatic information available for a given sample by fitting the   spectral energy distributions (SEDs).    The first step is to model the stellar  emission using  an evolutionary population synthesis method, assuming a star formation history  with some recipes for dust attenuation. We   then  compare these theoretical  SEDs to  data. This method  is particularly convenient  when a large range of redshift is studied and when the wavelength coverage is wide. Without  strong constraints on the amount of dust attenuation, an age-extinction degeneracy cannot be avoided (\cite{pforr12,conroy13}).  In parallel to the development of models for the stellar emission, IR ($> \sim 5 \mu$m) SEDs resulting from  dust  emission have been built and  models have been developed that predict the full UV to IR SEDs of galaxies in a self-consistent manner  for galaxies where most of the energy is produced by stars  (\cite{dacunha08,noll09}). The coupling of  emissions both from stars and dust is physically due to the absorption of UV-optical photons by dust. The exact process of dust and star interaction is highly complex  as it depends on many  parameters, such as the dust grain composition and distributions, geometry, and age of the stellar populations (\cite{popescu11}). At  the scale of entire galaxies, it is impossible to model this process and the net  effect of dust attenuation on the emission of a galaxy is usually described by an attenuation curve, the most popular one being that of \cite{calzetti00}. 
The $M_{\rm star}$ derivations  are not sensitive to dust attenuation,  but strongly depend on the assumed  stellar population synthesis model and  star formation history  (e.g., \cite{papovich01,salim07,pforr12,maraston10,lee09}). For a given set of assumptions about the stellar population synthesis model, including metallicity  and star formation history,  and adopting an   initial mass function,  the stellar masses can be  estimated robustly. However, systematic differences appear when different assumptions are made and these input parameters  are  very poorly  constrained (\cite{muzzin09}). \cite{marchesini09} showed that these systematic uncertainties contribute at the same level as random errors in the derivation of  stellar mass functions.   The uncertainty about  star formation history itself  induces large variations in stellar mass derivation  (\cite{bell03, lee09,pforr12}): the  high luminosity of the young stellar population may hide an old stellar population. The effect can be very strong in bursty systems and  leads to an underestimation of stellar ages \cite{maraston10}. As a consequence, stellar masses are likely to be more reliably measured in quiescent systems than in bursty galaxies (\cite{wuyts09, pforr12,conroy13}). \\

 In this paper, we aim to measure the SFR and $M_{\rm star}$ in a consistent way   using  different assumptions about  star formation history and  different  wavelength coverages.  We focus on   galaxies with a redshift larger than 1, which are   intensively  forming stars. Our analysis is based on a unique stellar population synthesis model (\cite{maraston05}). We refer to \cite{conroy10} for a comparison of the most popular population synthesis models in the framework of SED modelling. 
Numerous papers have been  published to explore the reliability of SED fitting techniques, most of which are  based on artificial catalogues of galaxies drawn from semi-analytical models or hydrodynamical codes (\cite{wuyts09,lee09,pforr12,pacifici12,mitchell13}). These studies are  all  based on UV-optical and NIR data,  but  do not include  IR emission. 
Our approach is somewhat different and complementary to these previous studies. First, we  consider  the whole electromagnetic spectrum from the UV to the IR and 
 we estimate dust attenuation by balancing the energy absorbed by the dust in the UV/optical to the energy re-emitted in the thermal infrared. The strong constraints put on dust attenuation are expected to reduce the uncertainty of recent star formation history retrieved by the SED fitting process. We  combine true data and mock catalogues built with our fitting code. The priors of our  artificial sources  created  with the fitting code  are  well defined and completely known, and their influence can be analysed at the price of  over-simplistic  modelling. Models based on semi-analytical codes  or hydrodynamical simulations are certainly more realistic, but also depend on the assumptions made to produce them and it is more difficult to quantify the influence of these priors on the results of the SED fitting analysis. \\

 We work between  z=1 and z= 3 in order to  sample the UV continuum and the dust emission well. All  galaxies are detected in the  IR  and, in particular, at  24 $\mu$m.  The choice of the redshift range is essentially motivated by the quality of the photometric data and the performance of the SED fitting code as described in Sect. 2 and 3.1.  The dataset is described in detail  in Sect. 2, with a    particular attention paid to  the  sampling of the rest-frame UV continuum emission  using a large number of intermediate band filters.
These filters have been proven to be very effective in measuring
photometric redshifts (\cite{ilbert09,benitez09,cardamone10})  as well
as in  characterising the dust attenuation (\cite{buat11,buat12}) but, to our knowledge, their influence in retrieving physical parameters has not been studied yet.  The fitting process, based on  our modelling code, CIGALE (\cite{noll09}), is presented in Sect. 3 along with the adopted  star formation models  and  the determination of the main physical parameters (SFR,  $M_{\rm star}$, and stellar ages). The influence  of  specific datasets (UV sampling, NIR, and IR)  in the determination of these quantities are explored in Sect. 4. In Sect. 5 we generate mock catalogues
and use them to explore possible systematic effects in the estimations of
the physical parameters. Finally, our conclusions are presented in 
Sect. 6.  We assume that $\Omega_{\rm m} = 0.3$, $\Omega_{\Lambda} = 0.7$, and $H_0 = 70\,{\rm km\,s^{-1}\,Mpc^{-1}}$. The luminosities are defined in solar units with ($L_\odot = 3.83~ 10^{33} ~ \rm erg s^{-1}$   and  the adopted  solar luminosity in the K band used to define $M/L_{\rm K}$ ratios is 5.13 $\rm 10^{32} erg s^{-1} Hz^{-1}$    (corresponding to  $M_{\rm K} = 5.19  ~{\rm ABmag})$.\\

\section{Observations and sample used}

The Great Observatories Origins Deep Survey Southern (GOODS-S)  field is  among the best observed fields for the purpose of cosmological studies. Its wavelength coverage is exceptional, combining photometric observations from the UV to the IR \footnote{http://www.stsci.edu/science/goods/} and spectral surveys to measure as many redshifts as possible \cite  [and reference therein]{kurk13}.  We select galaxies in this field with accurate measurements of the UV, visible, NIR, and IR rest frame emissions.\\
In the framework of the MUSYC project, \cite{cardamone10} compiled a uniform catalogue of multi-wavelength  photometry for sources in GOODS-S, incorporating the GOODS {\it Spitzer} IRAC and MIPS data (\cite{dickinson03}). In addition to broadband optical data, they used deep intermediate-band (IB)  imaging from the Subaru telescope to provide photometry with fine wavelength sampling and to  estimate more accurate photometric redshifts. In a previous work, we used  these data 
to trace  the detailed shape of the UV rest-frame continuum and to perform an accurate measurement  of  the dust attenuation curve  (\cite{buat11,buat12}). As mentioned by \cite{cardamone10} ,  the fluxes of extended sources  may be  underestimated in the  MUSYC catalogue. The reason is that total fluxes are deduced from aperture fluxes using the SExtractor's AUTO fluxes and a correction based on point source (stellar) growth curves. For extended sources this correction  underestimates  the total flux  by a factor that depends on both the size and magnitude of the sources  \cite[as shown in figure 7 of]  {taylor09}. Our selection of sources with $z > 1$ ensures that we  avoid this potential problem.
\\
We restrict the field  to the $10' \times 10'$  observed by the PACS instrument (\cite{poglitsch10}) on board  the $Herschel$ Space Observatory (\cite{pilbratt10}) at 100 and 160\,$\mu$m, 
as part of the GOODS-$Herschel$ key programme (\cite{elbaz11}).  The GOODS-$Herschel$ catalogue  is obtained from source extraction on the PACS images performed at the prior position of ${\it Spitzer}$ 24 $\mu$m sources, as described in \cite{elbaz11} and in the documentation provided with the data release \footnote{http://hedam.lam.fr/GOODS-Herschel}.

\begin{figure}
   \centering
  \includegraphics[width=\columnwidth]{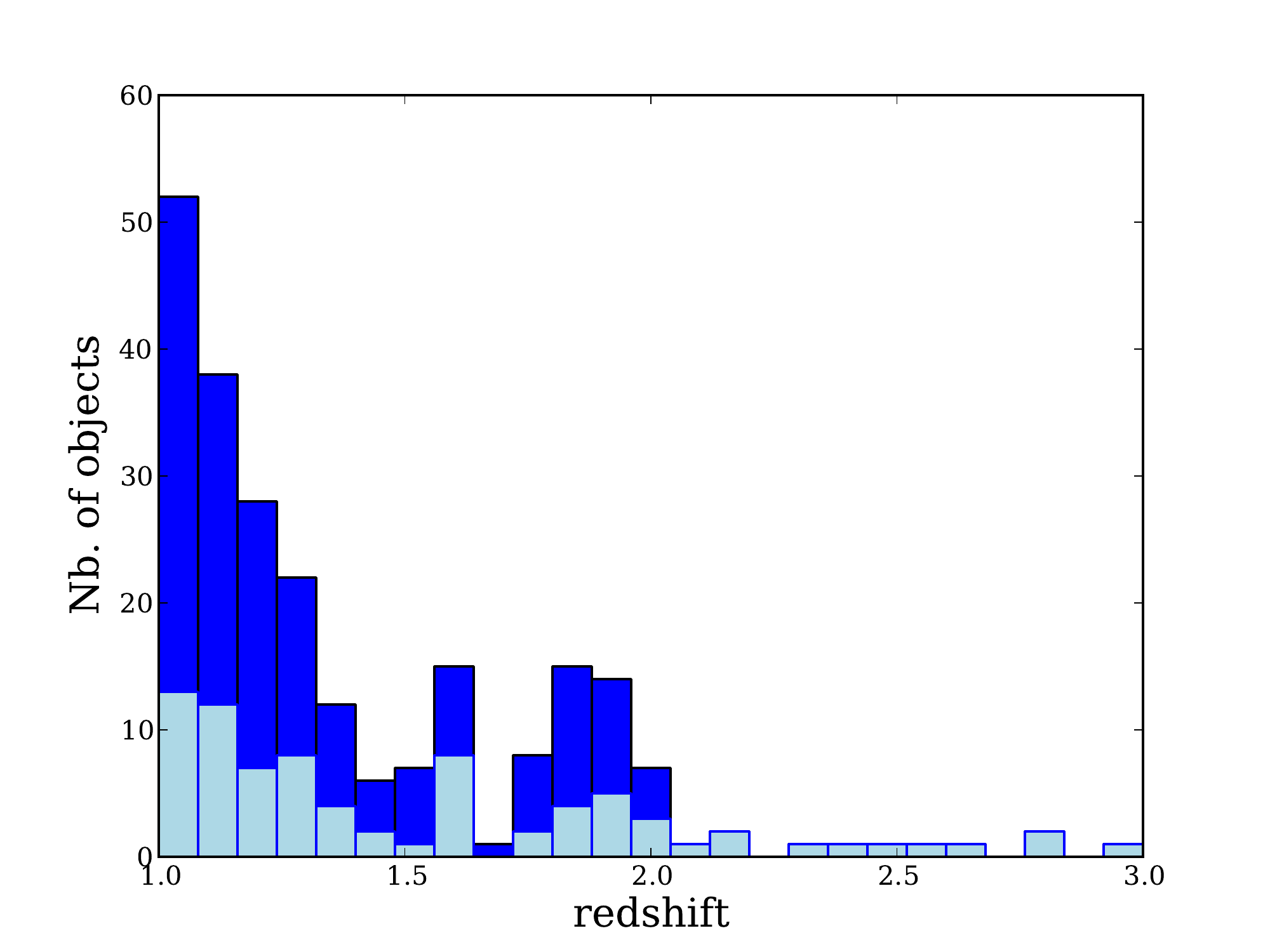}
   \caption{Redshift distribution of the sources. The light blue histogram is for sources detected in at least one PACS filter.}
              \label{zdist}%
    \end{figure}

We start with the sources detected at 24 $\mu$m with a S/N ratio larger than
3 and  we adopt the  Spitzer/IRAC detections used to extract the
24 $\mu$m sources (\cite{elbaz11}). These sources are cross-correlated with the MUSYC catalogue with
a tolerance radius between the IRAC and optical coordinates equal to 1
arcsec.  We further restrict the sample to sources that  are not detected
in the MUSYC X-ray catalogue of  \cite{cardamone08}. Our field corresponds  to the very deep $Chandra$ Deep Field-South survey reaching a limiting flux of  $\rm 1.7 10^{-16} erg s^{-1} cm^{-2}$ in the 0.5-2 keV band (CDFS-S A03). We also select sources such that they have  a single optical/UV counterpart within 3 arcsec.  Particular care is given to the redshift of the sources:  we select only galaxies of the MUSYC catalogue with a reliable spectroscopic redshift. All the  spectra were taken with VLT spectrometers FORS and VIMOS. The quality of the redshift  corresponds to more than 60$\%$ confidence level for the VIMOS surveys and labelled to be  'high quality, secure, and  likely' determinations for the FORS data (more detail can be found in the readme of the MUSYC catalogue and reference therein). As in Buat et al. (2012) (hereafter Paper I), we  consider only sources with  at least two measurements at $\lambda<1800\AA$ in the rest frame of galaxies (broad or IB bands ) with a S/N ratio larger than 5. This condition ensures that we have a good definition of the UV continuum in the wavelength range mostly used to measure star formation rates ($1500-1600 \AA$) and for which  calibrations are available in the literature.
We obtain 312 galaxies that satisfy  all  these conditions.  
To combine the PACS observations with 
this sample we need to apply de-blending techniques to measure  fluxes
because of the large size of the PACS beam (\cite{elbaz11}). We follow the prescription given in the documentation provided with the data release and  only consider PACS measurements (with S/N > 3) for sources without bright neighbours defined as   being brighter than half the flux density of the source and closer than 1.1 of the FWHM of the PSF. We add a further condition that no 24 $\mu$m bright neighbour must be found close to the  100 and 160 $\mu$m detections (i.e. no  source brighter than half the flux density of the source  at 24 $\mu$m and closer than  1.1 of the full width half maximum (FWHM)  of the point spread function (PSF) at 100 or 160 $\mu$m). In the end,   we obtain 100 $\mu$m fluxes for 92 out of the 312  sources and  160 $\mu$m fluxes for 54.  Finally, for z $>$  2, we only keep the sources detected with PACS since the 24 $\mu$m data correspond to rest frame wavelength lower than 8 $\mu$m and are not considered reliable to measure total IR emission from dust. With only two sources at $z > 3$, we reduce  the sample to the 236 sources with $1<z<3$.   We detect 32 objects  with PACS at 100 and 160 $\mu$m, 40 only at 100 $\mu$m,  and eight only at 160 $\mu$m. The redshift distribution is shown in Fig.\ref{zdist}. We compile 28 photometric bands  from U to 24 $\mu$m: the optical and NIR broadbands ($U, U38, B, V, R, I, z, J, H, K$),  13 IB bands from 427 to 738 nm, the four IRAC bands,   and the MIPS1 24 $\mu$m band.  We apply the revised IRAC selection criteria of \cite{donley12} to check that the 230 galaxies detected in the four IRAC bands all fall out of the AGN selection region defined in the $f_{5.8}/f_{3.6}$ and $f_{8}/f_{4.5}$ colour plot.
  The rest-frame luminosity at 1530\,$\AA$ (corresponding to the FUV $GALEX$ filter) of each galaxy is obtained by modelling a powerlaw, $f_\lambda ({\rm erg\,cm^{-2}\,s^{-1}\,\AA^{-1}}) \propto \lambda^{\beta}$ between 1300 and  2500 $\AA$,  in the rest frame of the sources. In the following, we define $L_{\rm FUV}$ as $\nu L_{\nu}$ at 1530\,$\AA$. 

\begin{table*}
\caption{Input parameters for SED fitting with CIGALE, in boldface parameters used to create artificial galaxies (Section 6) }
\label{tab:parameters}
\centering
\begin{tabular}{c c c}
\hline
Parameter & Symbol & Range \\
\hline\hline
Amount of dust attenuation  \tablefootmark{1} & $A_V$ &{\bf  0.25, 0.5, 0.75}, 0.9,{\bf 1.05}, 1.2, 1.35, {\bf 1.5}, 1.65, {\bf 1.8}\,mag \\
\hline
\hline
Two stellar populations \tablefootmark{2}, age  free\\
\hline
age of the stellar population & $t_f$ & {\bf 1, 2, 3, 4, 5}\,Gyr \\
$e$-folding rate of the old stellar population & $\tau$ & {\bf 1, 3}\,Gyr \\
age of the young stellar population & $t_{\rm ySP}$ & {\bf 0.01, 0.03, 0.1, 0.3}\,Gyr \\
stellar mass fraction due to the young stellar population & $f_{\rm ySP}$ & {\bf 0.01, 0.02, 0.05, 0.1, 0.2}, 0.5 \\
\hline
\hline
One  stellar population with an exponential SFR, age  free\\
\hline
age of the  stellar population & $t_f$ & 0.05,0.1,0.25, 0.5,1,2,3, 4, 5\,Gyr \\
$e$-folding rate of the stellar population & $\tau_1$ & 0.5,1, 3, 5, 10\,Gyr \tablefootmark{3} \\
\hline
\hline
Models with  a fixed age for the stellar population   \tablefootmark{4}\\
\hline
$1<z<1.2$& $t_f$& 5 Gyr\\
$1.2\le z\le 1.5$& $t_f$& 4 Gyr\\
$1.5\le z \le 2$& $t_f$& 3 Gyr\\
$2\le z \le 3$& $t_f$& 2 Gyr\\
\hline
\end{tabular}
\tablefoot{Values of input parameters used for the SED fitting with CIGALE. Bold face values are used for the mock catalogue in Section 5. \\
\tablefootmark{1}{In case of two populations, $A_V$ corresponds to the attenuation of the youngest population and   a reduction factor of 0.5 is applied to the visual attenuation of the older stellar population (see text) }\\
\tablefootmark{2} { One  exponentially decreasing stellar population with a younger stellar component with a  constant  SFR}\\
\tablefootmark{3}{Negative values of $\tau_1$ for exponentially increasing star formation rates}\\
\tablefootmark{4}{In case of two stellar populations, only the age of the oldest stellar population is fixed}\\

}
\end{table*}

\section{SED fitting}

The SED fitting is performed with the CIGALE code (Code Investigating GALaxy Emission)\footnote{http://cigale.lam.fr} developed by \cite{noll09}.  The code CIGALE combines a UV-optical stellar SED with a dust component emitting in the IR and fully conserves the energy balance between the dust absorbed stellar emission and its re-emission in the IR. In the present work, we do not  perform as detailed a study of dust attenuation as in Paper I. Instead, we  keep  the parameters of the attenuation curve (amplitude of the $2175 \AA$ bump and slope of the UV attenuation curve) free since dust attenuation curves are expected to vary among galaxies (\cite{witt00,inoue06}). We have checked that the estimated output parameters are fully consistent with the results obtained in Paper I.  We adopt the stellar populations synthesis models of \cite{maraston05}, a Kroupa (\cite{kroupa01}) initial mass function (IMF),  and a solar metallicity as our baseline. Different star formation histories are considered as described below.\\
Dust luminosities (L$_{\rm IR}$ between 8 and 1000\,$\mu$m) are computed by fitting \cite{dale02} templates and are linked to the attenuated stellar population models: the stellar luminosity absorbed by the dust is re-emitted in the IR. The validity of \cite{dale02} templates for measuring total IR luminosities of sources detected by $Herschel$ is confirmed by the studies of \cite{elbaz10,elbaz11}. A single parameter $\alpha$ describes these templates, defined as the exponent of the distribution of dust mass over heating intensity.  When the source is detected in at least one PACS band, the input values of $\alpha$ are 1, 1.5, 2, and 2.5. As in Paper I, when PACS data are not available $\alpha$ is assumed to be  2, which corresponds to  the average value found for galaxies detected with PACS. We have checked that the predicted fluxes  in the PACS bands  are consistent with a non-detection of these galaxies with PACS at a 3$\sigma$ level quoted by  \cite{elbaz11}.
The input parameter measuring dust attenuation  is  the attenuation in the V band (for the young stellar population if two stellar populations are involved), the input values used in this work are listed in Table 1. Dust attenuation  in the FUV band is  also defined as an output parameter of the code.   \\
The parameters  are all estimated from their probability distribution function (PDF) with  the expectation value and its standard deviation, it is described as the 'sum' method in \cite{noll09} and \cite{giovannoli11}.    
In addition to the input parameters, the output parameters considered in this work will be   $M_{\rm star}$,  SFR, instantaneous and averaged over 100 Myr,  ages of the stellar populations, and dust attenuation in the FUV ($A_{\rm FUV}$). 

\subsection{Star formation histories} 
We use different star formation histories  (SFHs) since we want to test their influence on parameter estimations. All the input parameters related to the SFH are listed in Table 1. Three  scenarios are implemented in CIGALE and we consider all of them.  Models are built  with an age of  creation of the first stars  that can be either a   free or fixed parameter. To fix the age of the stellar population we  follow \cite{maraston10} (see also \cite{pforr12}) and  adopt a redshift formation $z_f \simeq 8$. Practically the whole redshift range is split to four   intervals with different age models corresponding to a redshift formation of 8 for the central redshift of each interval (Table 1).   The different SFH considered in this work are:
\begin{enumerate}
      \item    A single stellar  population  with  exponentially decreasing SFR and an e-folding rate $\tau_1$, called hereafter decl.-$\tau$ model. The age $t_f$ of the stellar population  is left a free  parameter since adopting a fixed age would be  unrealistic for active star forming galaxies and would not produce reliable results. This model is still  commonly used in the literature (\cite{ilbert13, muzzin13}) although it has long been identified to be unrealistic (\cite{boselli01}), leading to very young ages for the stellar populations (\cite{maraston10}).
        \item  A single stellar  population  with  exponentially increasing SFR, called hereafter  rising-$\tau$ model.  The parameters are identical to those for the decl.-$\tau$ model, but with the age of the stellar population left  either free or fixed. Several recent studies of distant galaxies prefer similar scenarios with rising SFR  (\cite{maraston10,papovich11,pforr12,reddy12}).
      \item  Two stellar populations:  a recent stellar population with a constant SFR on top of an older stellar population created with an exponentially declining SFR. The parameter $t_f$ is the age of the older stellar population, it is  chosen to be either free or fixed. The age $t_{\rm ySP}$ of the young component is always a free parameter. The two stellar components are linked by their mass fraction $f_{\rm ySP}$. Such models are introduced since they are expected to better reproduce real systems, which experience several phases of star formation  (\cite{papovich01, borch06,gawiser07, lee09}).
   \end{enumerate}
The complete  grid of values  is used to build the PDFs. In what follows we sometimes refer to Decl.-$\tau$ and rising-$\tau$ models  as $\tau$-models. For the two-populations model, a reduction factor of the visual attenuation, $f_{\rm att}$, is applied to the old stellar population to account for the distributions of stars of different ages (\cite{charlot00, panuzzo07}). From our previous analyses, we adopt $f_{\rm att}=0.5$. The results are not sensitive to the exact value of this parameter.\\

More complex SFHs are  also proposed in the literature. Delayed SFR  ($ t \exp(-t/\tau$))  are not fully conclusive overall (\cite{lee10, lee11,schaerer13}). More realistic  histories derived from modelling of galaxy evolution  combining power law and exponential variations \cite{buat08,behroozi12}  are promising. They are not yet implemented in CIGALE, but will be in its future version (Burgarella et al. in preparation). The treatment of emission lines is not optimal in the version of the code used in here. As a result we do not consider sources with $z<1$ and those for which IB filters  enter the rest-frame optical range where emission lines can have a large impact (\cite{kriek11,schaerer13}).
\begin{figure}
   \centering
 \includegraphics[width=\columnwidth]{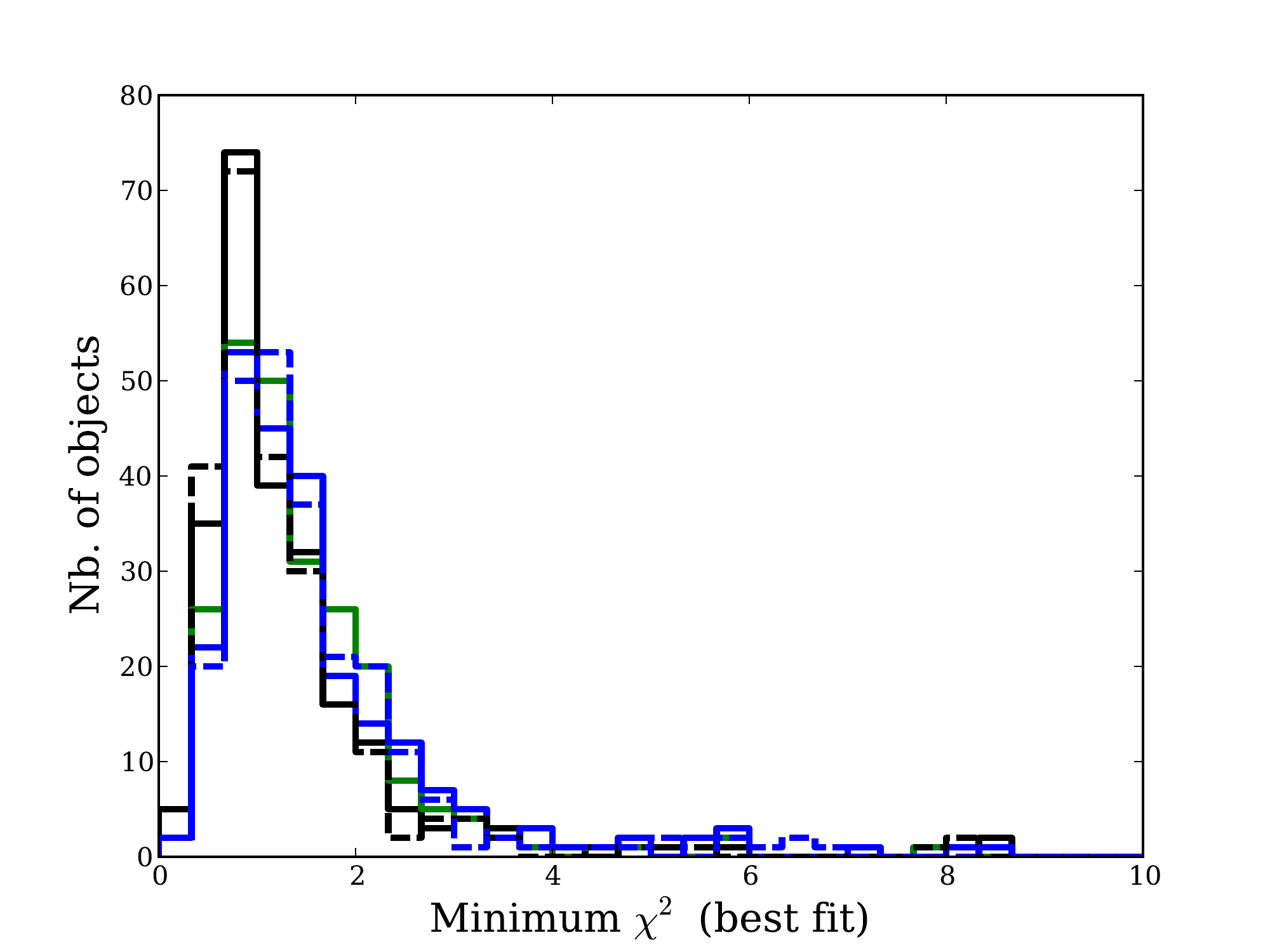}
   \caption{Distributions of $\chi^2$ values obtained for the best fits and plotted in a logarithmic scale. Black: two-populations model,  green: decl.-$\tau$ model, blue: rising-$\tau$ model. The solid lines are obtained for models  with  the age of the stellar population left free, the  dashed lines correspond to a fixed age for  the stellar population.  }
              \label{Khi2dist}%
    \end{figure}

\subsection{Results of the spectral energy distribution fitting process}
An initial, global, comparison of the three models of SFH can be performed by comparing the reduced best-fit $\chi^2$ distribution  given by the code  as shown in Fig. 2. Note that the derived physical parameters are not directly retrieved from the best model,  but from the analysis of their probability distribution function calculated with all the input models. The number of degrees of freedom to calculate the reduced $\chi^2$ varies from 4 for  the fixed-age $\tau$-models to 7 for the free-age two-populations model.  The best-fit $\chi^2$ distributions appear  similar   for all models.  Best-fit $\chi^2$ values lower than 3 are found  for more than  $\sim 90 \%$ of the sample with median value of $\chi^2$ equal to 1 for both  fixed-age and  free-age  two-populations models and 1.3 for the one-population models (age-free decl.$\tau$ and age free and fixed rising.$\tau$ models).  We conclude that the models considered  reproduce  our data  satisfactorily with a slightly better fit obtained with the two-populations models. 

\subsubsection{Stellar ages}
\begin{figure}
   \centering
  \includegraphics[width=\columnwidth]{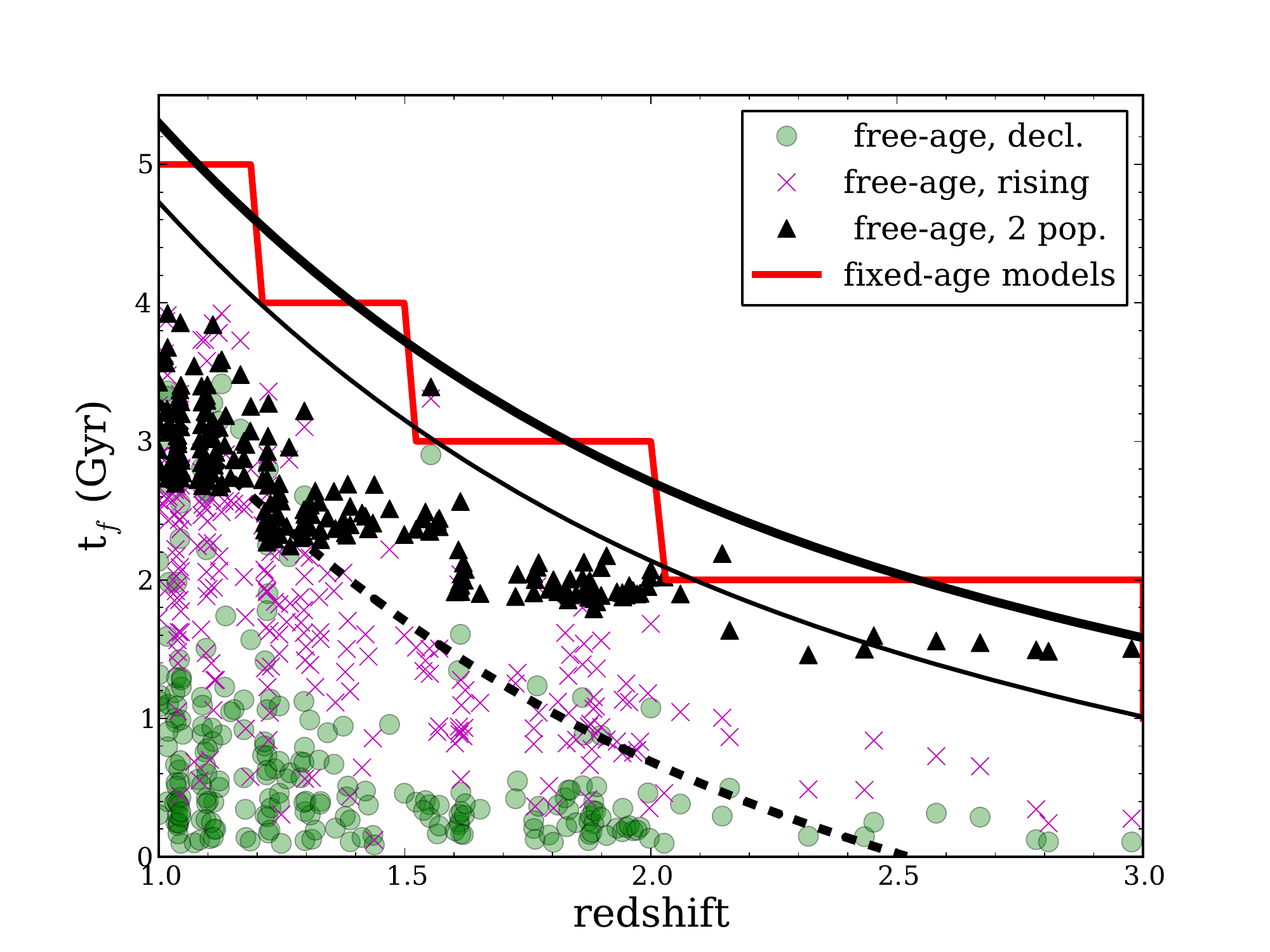}
    \includegraphics[width=\columnwidth]{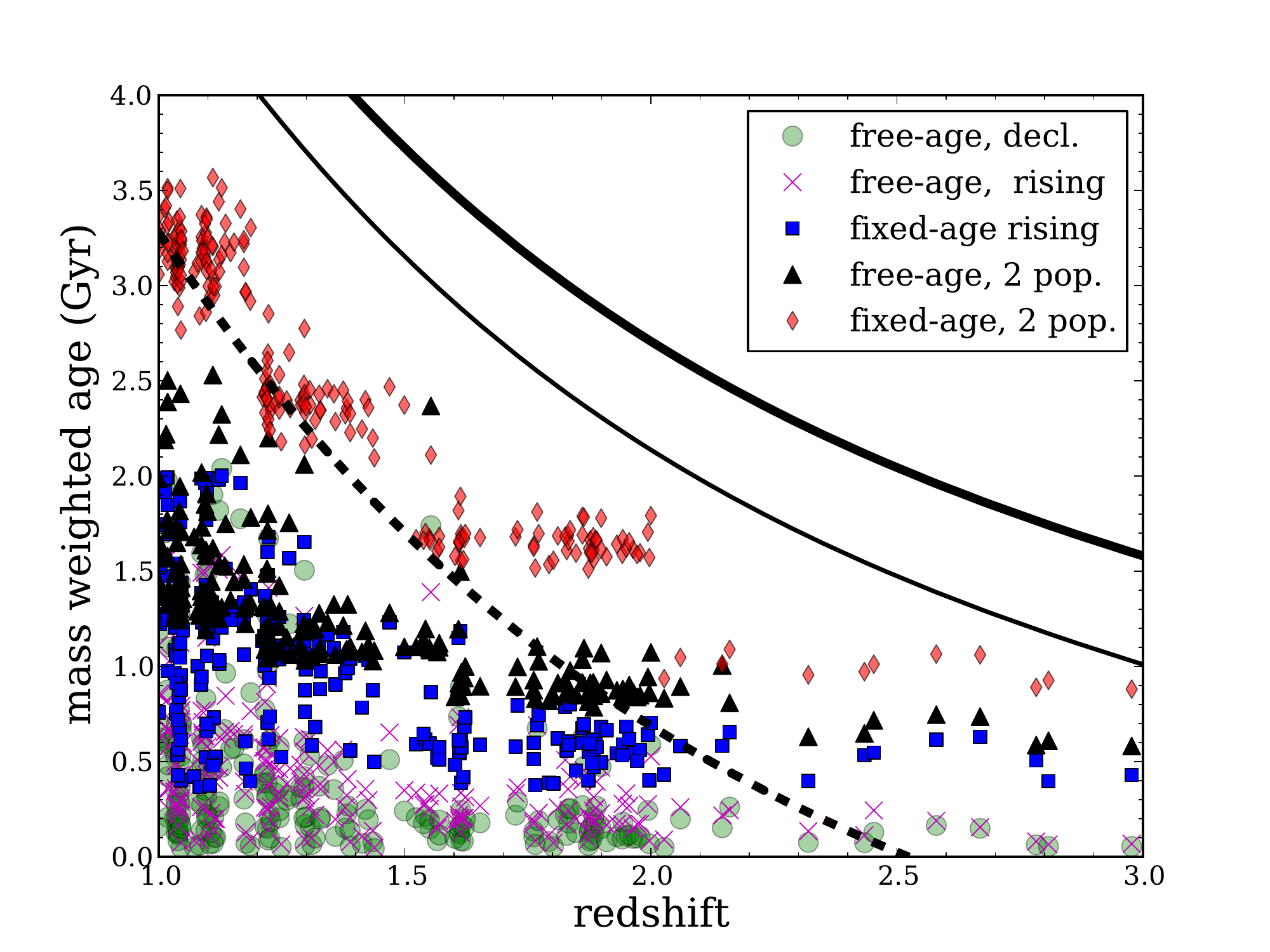}
   \caption{Stellar ages estimated with CIGALE. Upper panel: ages of the (oldest) stellar population for free-age  models as a function of redshift. Black triangles: two-populations model, green circles:  decl.-$\tau$ model,  magenta crosses: rising-$\tau$ model. The red steps correspond to the age adopted for the models with fixed ages. Lower panel: mass weighted ages for all the models: same symbols as in the upper panel  are used for the free-age models, blue squares  and red diamonds are  added for the fixed-age rising-$\tau$ model and two-populations models, respectively.  The  curves correspond to the age at  redshift z  for  a   redshift formation $z_f =8$ (thick line), $z_f=5$  (thin line), and $z_f =2.5$ (dotted line). }
              \label{z-age}%
    \end{figure}

 In the upper panel of Fig. \ref{z-age} we   compare the  ages  $t_f$ when  star formation begins, obtained for each  model considered.   The estimated ages are reported with the ones adopted  for the fixed-age models. Very young and quite unrealistic ages  for the beginning of the star formation are found with the  free-age decl.-$\tau$ model. With the free-age  rising-$\tau$ model the estimated ages are higher, yet remain quite low. It is an illustration of the well-known outshining of the young populations in galaxies forming stars actively as nicely illustrated by \cite{maraston10} (their Fig. 12) .
 Ages obtained with the free-age two-populations  model are more  realistic: combining a several billions-years-old population with a much more recent  one ($< 300$ Myr) gives more flexibility  to 
 account for the main  light production by young stars with an underlying older population dominating the mass. 
    The ages of the old population found with the free-age two-populations model correspond to a redshift formation from 2.5 (for galaxies  at $z=1$) to five (for galaxies at $z=2$).  If we assume that we are observing the same population of galaxies  from $z=2-3$ to $z=1$, the  age of the stellar populations  found at different redshifts must be consistent: it  is not the case for our free-age  models. 
  However, we   may observe different galaxies at $z=1$ and $z=2$  since  galaxies active in star formation at $z=2$ may become quiescent at $z=1$. Combining the average SFH \cite{heinis13b} derive from the SFR-$M_{\rm star}$ relation at different redshifts and  assuming that galaxies exit the main sequence when they reach a mass characteristic of quiescent systems at the corresponding redshift,  \cite{heinis13b} estimate the time galaxies can stay on the main sequence. Galaxies with $M_{\rm star} > 10^{10} M_{\odot}$ (which correspond to the bulk of our sample as shown in Section 3.2.2) located on the main sequence at $z\ge 2$ reach their quenching mass before  $z=1$ and may well have evolved out of the main sequence . \cite{heinis13b} also find that   time spent on the main sequence  decreases when $M_{\rm star}$ increases. \\
  
The mass weighted age is also an output parameter of CIGALE. This age is representative of the average age of  stellar populations.  The values  are reported in the lower panel of Fig. \ref{z-age} as a function of redshift. The general trends are similar to those found in the upper panel with global shifts to lower ages. The free-age rising.$\tau$ and decl.-$\tau$ models lead  to similar weighted ages whereas $t_f$ values were found to be larger for the free-age  rising.$\tau$ models. For these models, the bulk of the stellar mass is built well after the beginning of the star formation. The fixed-age rising.$\tau$ model leads to slightly larger, but still very short, mass weighted ages.\\

The free-age decl.-$\tau$ models lead to very unrealistic ages, but we  keep them, because of their popularity. Concerning the rising-$\tau$ models, the free-age option also leads to very young ages. Rising-$\tau$ models are commonly used by adopting a fixed formation redshift, which allows a high redshift formation  with an average age for the bulk of the stellar population that remains young (\cite{maraston10,papovich11,pforr12}). Therefore, we decided to keep only the fixed-age rising-$\tau$ model. For the two-populations models, the free-age option leads to plausible ages as discussed above, therefore we will keep both free-age and fixed-age models.

For the purpose of comparison between all these models,  we adopt the free-age two-populations model  as our  baseline in the following.

 \subsubsection{Stellar mass and star formation rate  determinations}
 \paragraph{Variation  of the star formation history:}
 
\begin{figure}
   \centering
 
    \includegraphics[width=\columnwidth]{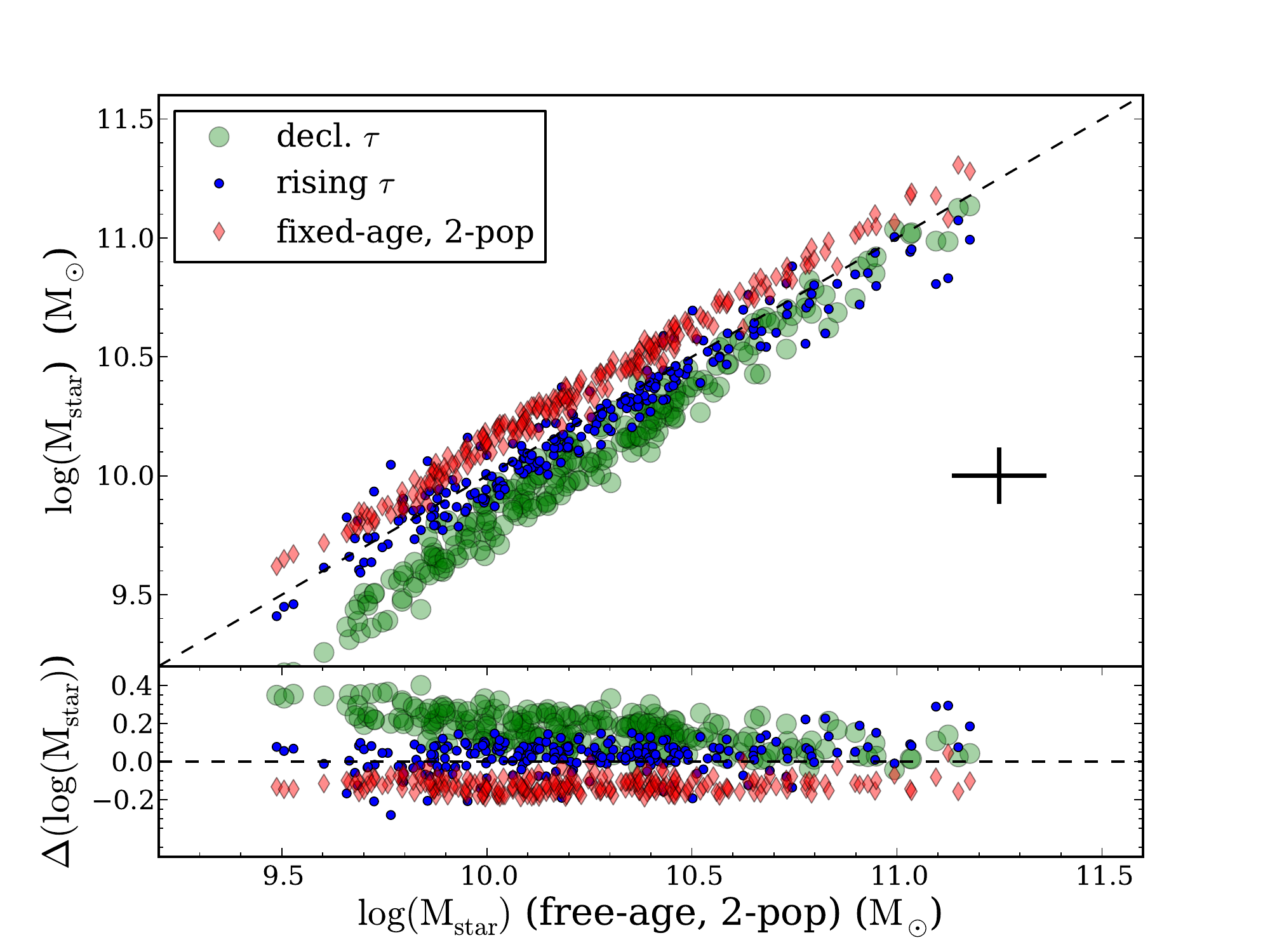}
      \includegraphics[width=\columnwidth]{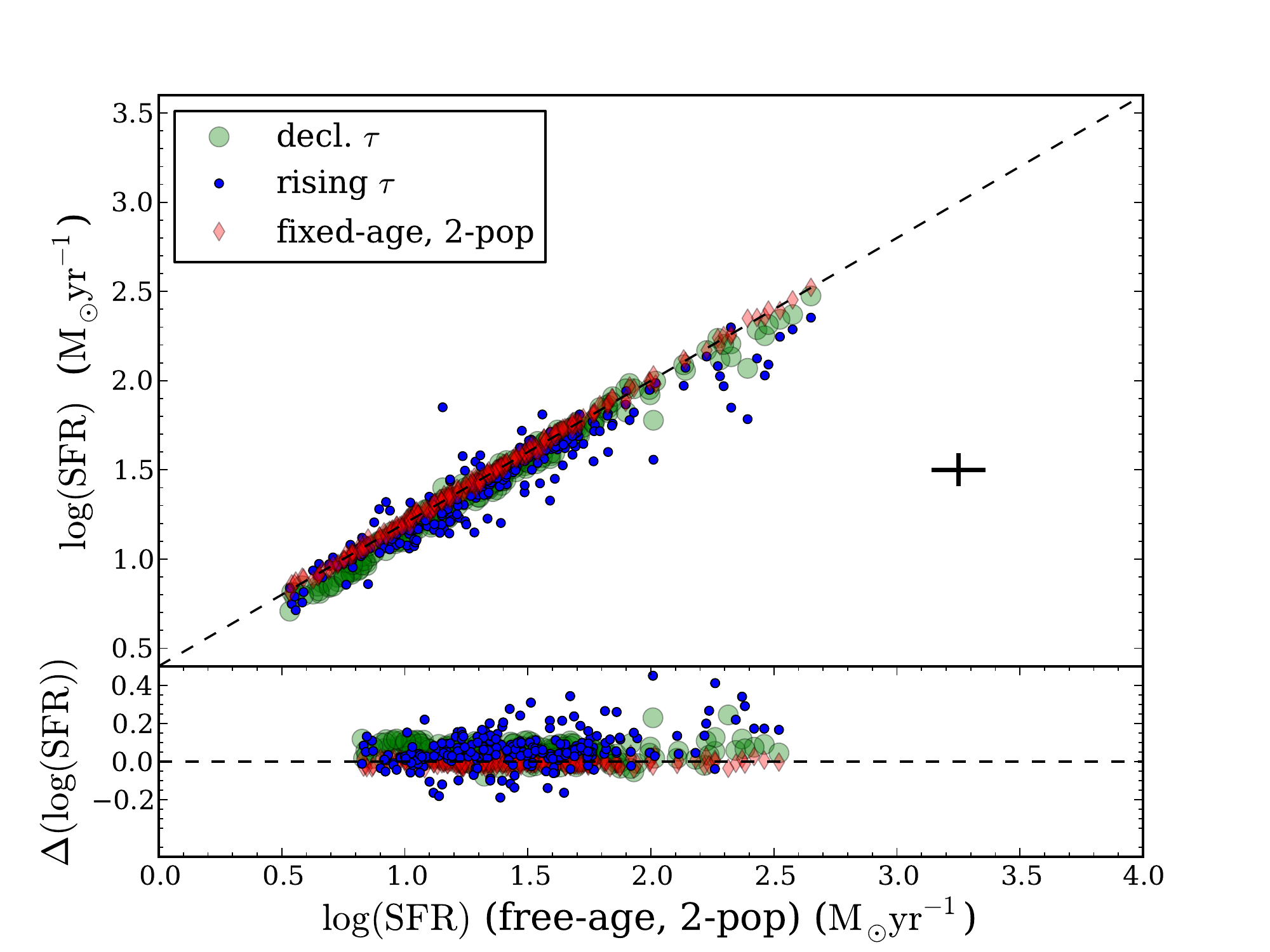}
   \caption{ Comparison of  $M_{\rm star}$ (upper panel) and SFR (lower panel) determinations  from the different    models. The x-axis is from the baseline model (free-age 2-populations), the y-axis corresponds to free-age decl.-$\tau$  (green circles), fixed-age rising-$\tau$ (blue dots),  and two-populations (red diamonds) models. Typical uncertainties on parameter estimations are indicated by a black cross. $\Delta(SFR) =    \Delta(\log(SFR_{\rm 2 pop})-\log(SFR_{\rm  \tau-model})$ and $\Delta(M_{\rm star} = \Delta(\log(M_{\rm star-2 pop} )-\log(M_{\rm star-rising-\tau }) ).$}
              \label{Mstars}%
    \end{figure}

SFR and $M_{\rm star}$ estimates  depend a priori on the assumed SFH. In this section, we consider the instantaneous SFR.\\
In Fig. \ref{Mstars},  we  compare $M_{\rm star}$ values  for all the models considered in this work, with our baseline model taken as the reference (x-axis). A good agreement is found with the rising-$\tau$ model with $ \Delta(\log(M_{\rm star-ref.mod} )-\log(M_{\rm star-rising-\tau }))~= 0.03\pm 0.08 $ dex. Substantial differences are found between the reference   model and  the same model with $t_f$ fixed:  fixing the age of the oldest stellar population  increases the stellar mass  by $0.12\pm 0.04$ dex. The lowest values of $M_{\rm star}$ are found for the free-age decl.-$\tau$ models with $\Delta(\log(M_{\rm star-ref.mod} )-\log(M_{\rm star-decl.-\tau }))~~= 0.17 \pm 0.09$ dex. The difference between the extreme cases (free-age decl.-$\tau$ and fixed-age two-populations models) reaches 0.3 dex.\\
SFR determinations are much more   consistent with each other. The agreement is almost perfect between both two-populations models. The star formation is dominated by the young stars formed in the more recent burst, which is fitted in the same way whether or not the age of the older population is free or fixed.  A very slight shift towards higher SFR  is found  for the two-populations models as compared to the two other ones:     $\Delta(\log(SFR_{\rm 2 pop})-\log(SFR_{\rm decl.-\tau})~=~0.04$ dex   and $\Delta(\log(SFR_{\rm 2 pop})-\log(SFR_{\rm rising-\tau})~=~0.06$ dex. \\

We confirm that the choice of the SFH  changes   the $M_{\rm star}$ determinations significantly (\cite{pforr12}). Because they are  outshined by young stars, older stellar populations are hidden and $\tau$- models are likely to  give   only lower limits to the total stellar mass (\cite{papovich01,pforr12}). Models considering an old and young population are known to  yield higher masses (\cite{papovich01,borch06,lee09,pozzetti07}). The measurement of  SFR  are found to be robust against changes in the SFH.  $\tau$ models yield results consistent with two-populations models with  a constant star formation for the recent burst. Actually,  \cite{reddy12} have shown that for ages larger than $\sim 100$ Myr rising-$\tau$ models  lead to a constant production of the UV light. In  the present study stellar populations are older than 2 Gyr for the rising-$\tau$  model, so we expect a robust determination of the SFR with the fit of the intrinsic UV continuum.   
 
The effective SFH  we obtain  for  the decl.-$\tau$ model   is close to  a constant SFR. The $t_{\rm f}/\tau$ ratio    is always  lower than unity  and the age of the stellar population is  larger than 100 Myr    for $98\%$ of our galaxies, ensuring a stationarity in the production of the UV light. Therefore we expected an agreement between our SFR estimations. The presence of IR data provides  a robust measurement of the dust attenuation, allowing us to  fit the true intrinsic UV continuum of our galaxies. The large impact of dust attenuation corrections will be discussed in Sect.  4 and 5.

\begin{figure}
   \centering

  \includegraphics[width=8cm,height=6cm]{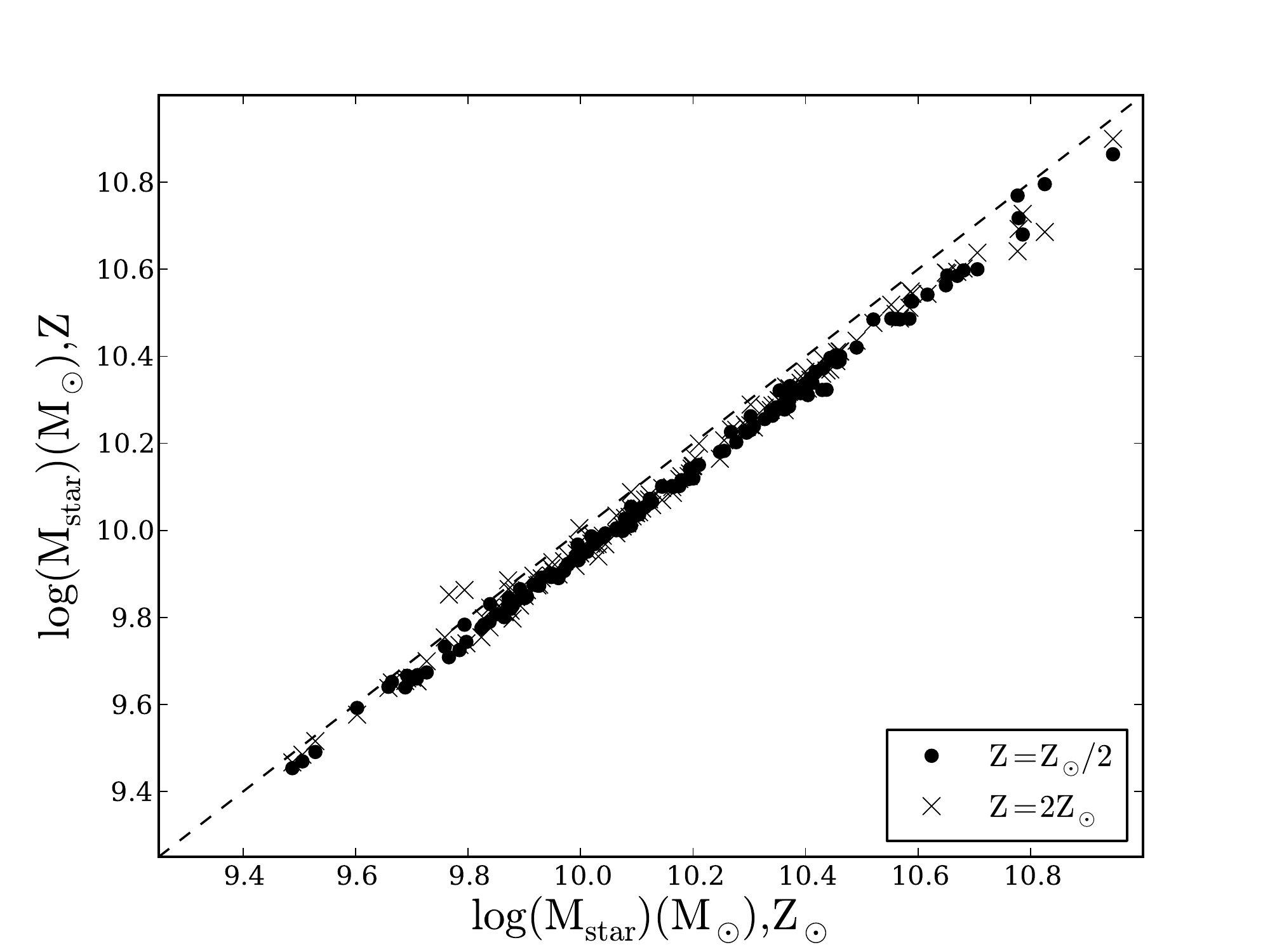}
    \includegraphics[width=8cm,height=6cm]{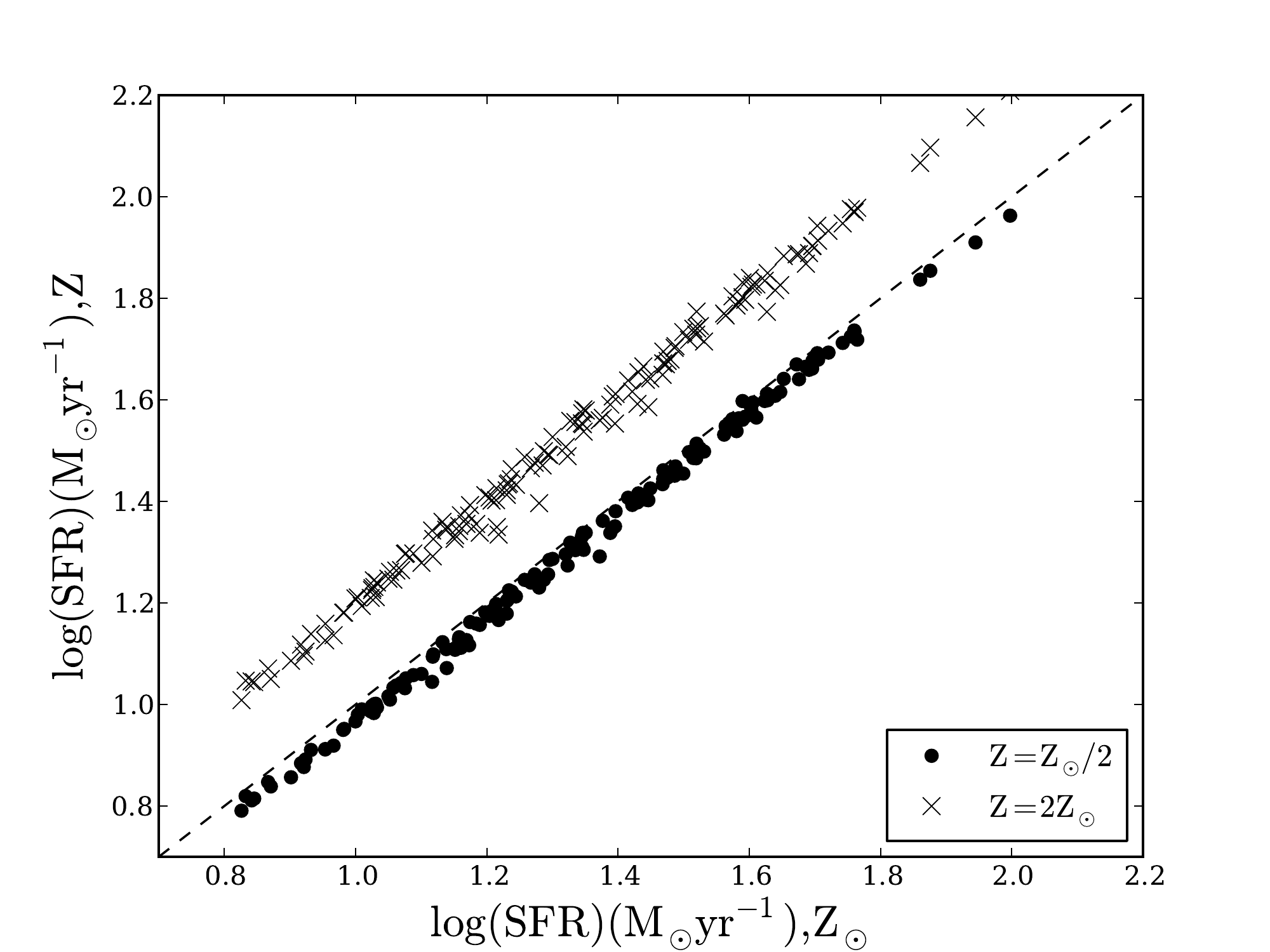}   
   \caption{ Comparison of  $M_{\rm star}$ (upper panel) and SFR (lower panel) determinations  for different metallicities. The quantities plotted on the x-axis are obtained with  the baseline model assuming a solar metallicity, the y-axis corresponds to determinations with a half solar (dots) and twice solar (crosses) metallicity.}
              \label{metal}%
    \end{figure}

 \paragraph{Variation of the IMF:}
Throughout this paper we adopt a Kroupa IMF. CIGALE allows us to use a Salpeter IMF (\cite{salp55}). Several authors have studied the influence of the choice of the IMF on the SFR and $M_{\rm star}$  determinations (\cite{bell03,brinchmann04,pforr12}). We confirm that changing the IMF from Kroupa to Salpeter increases  SFR and stellar masses  by a constant factor found equal to  $0.17\pm0.01$ dex and  $0.21\pm0.01$ dex respectively, independent of the adopted SFH. 
\paragraph{Variation of the metallicity:}
Throughout the paper we adopt a solar metallicity ($\rm Z_\odot$), but we can run CIGALE with Maraston models  with $\rm Z_\odot/2$ and $\rm 2 Z_\odot$. The comparison of the SFR and $M_{\rm star}$ determinations for these different metallicities are shown in Fig. \ref{metal}. The values obtained with non-solar metallicities correlate very well with those obtained with $\rm Z_\odot$. A very small systematic shift is found for $M_{\rm star}$ determinations, the stellar masses measured with  a non-solar metallicity being lower than the reference metallicity by $\sim 0.05$ dex. This  is likely due to the fact that stars in both sub-solar and super-solar models have slightly smaller turnoff masses (Maraston, private communication). The effect is larger for SFR measurements in the case of super-solar metallicity. The SFR is increased by 0.20 dex for $\rm 2 Z_\odot$ models (and decreased by only  0.03 dex for  $\rm Z_\odot/2$). This shift is attributed to a decrease of the UV photons for a given SFR in super-solar environment. 

  \begin{figure}
   \centering
       \includegraphics[width=\columnwidth]{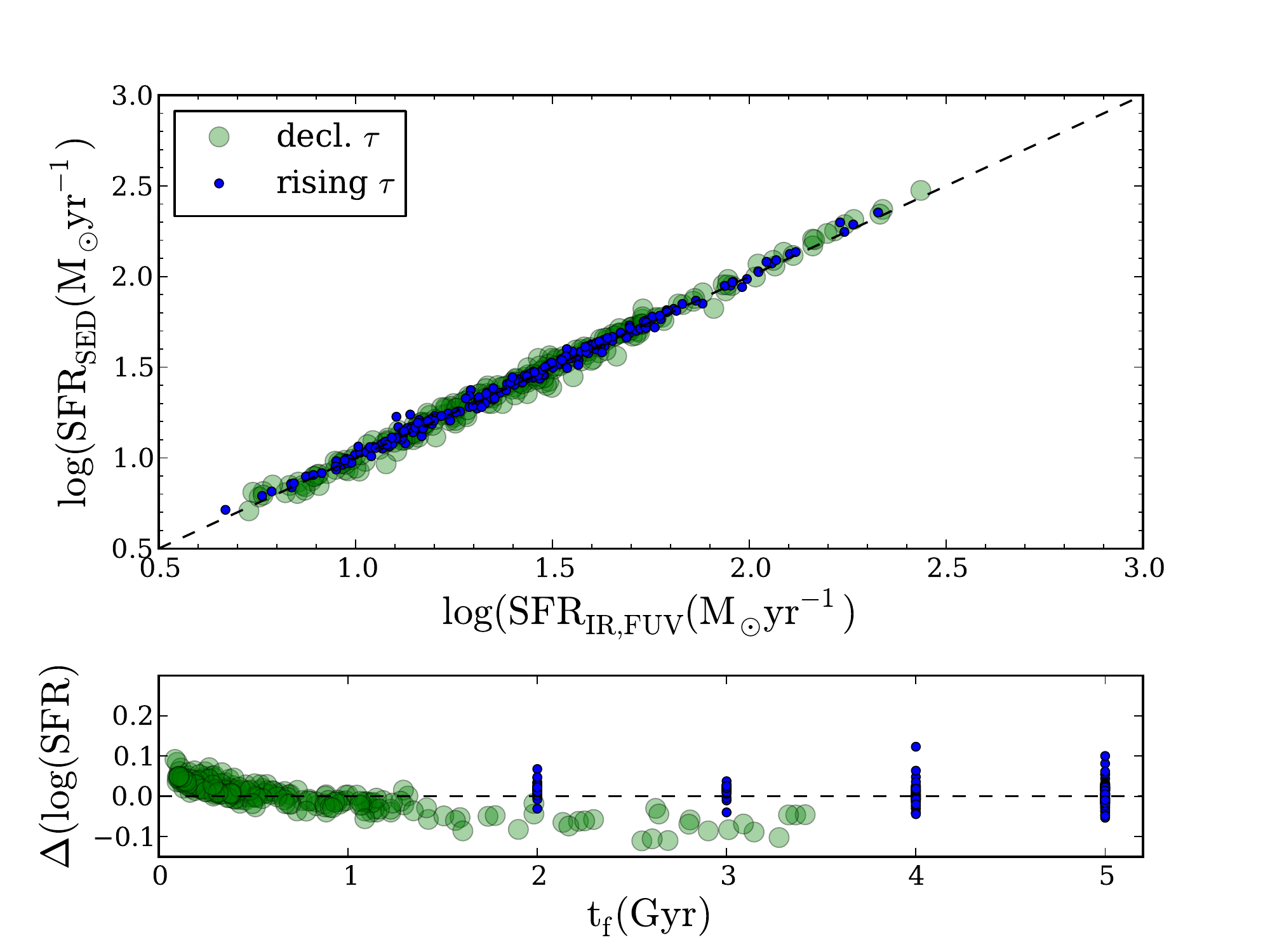}
      \includegraphics[width=\columnwidth]{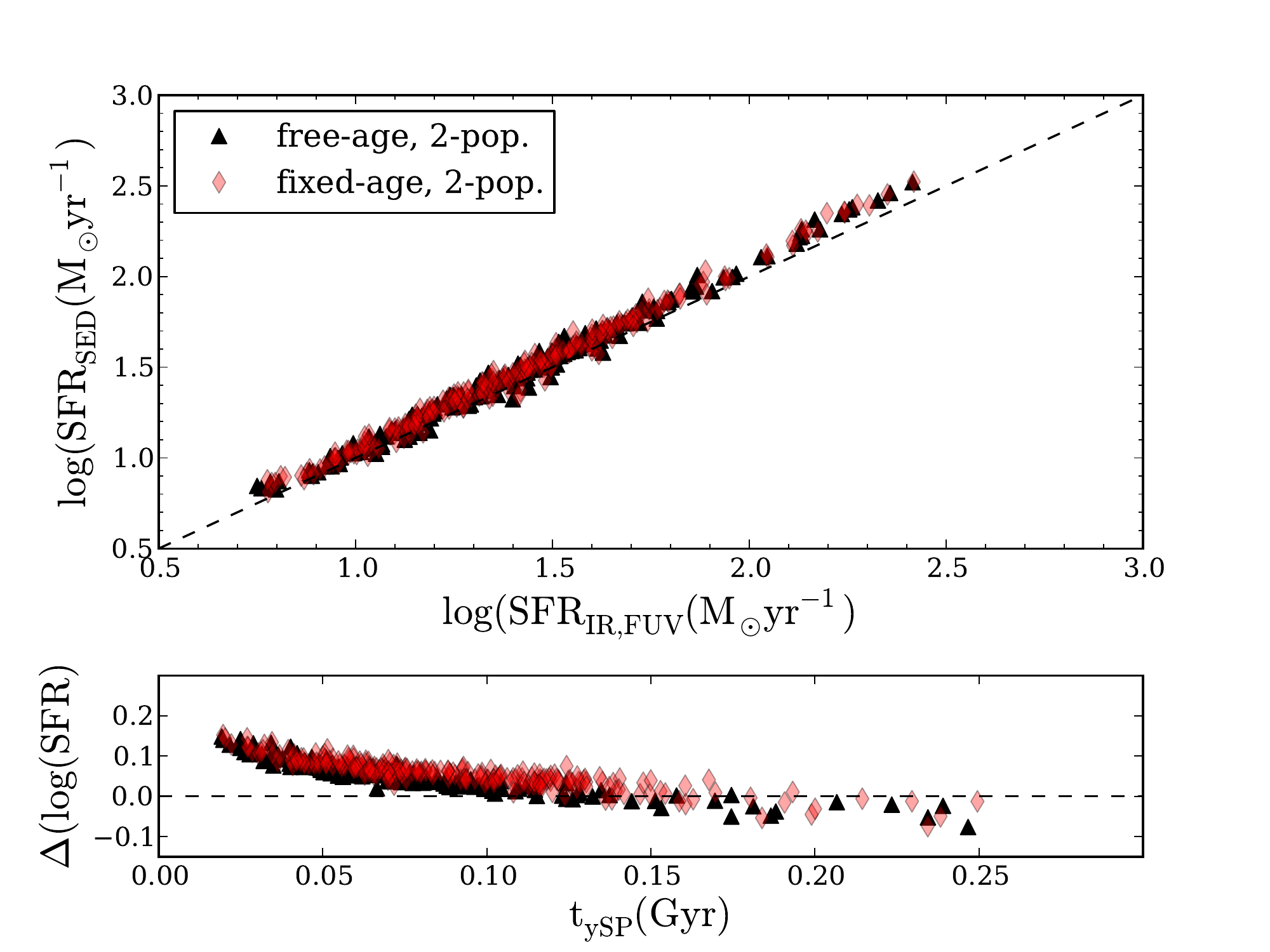}
   \caption{Comparison between $SFR_{\rm IR,FUV}$  (x-axis) and $SFR_{\rm SED}$ (y-axis): the two upper panels are   for   $\tau$-models , the two lower panels for   the two-populations model,   the  difference between $SFR_{\rm SED}$ and $SFR_{\rm IR,FUV }$ is  plotted against the age of the  (youngest) stellar population. Symbols are the same as in Fig. \ref{z-age}.}
              \label{comp-SFR}%
    \end{figure}
  \begin{figure}
   \centering
\includegraphics[width=\columnwidth]{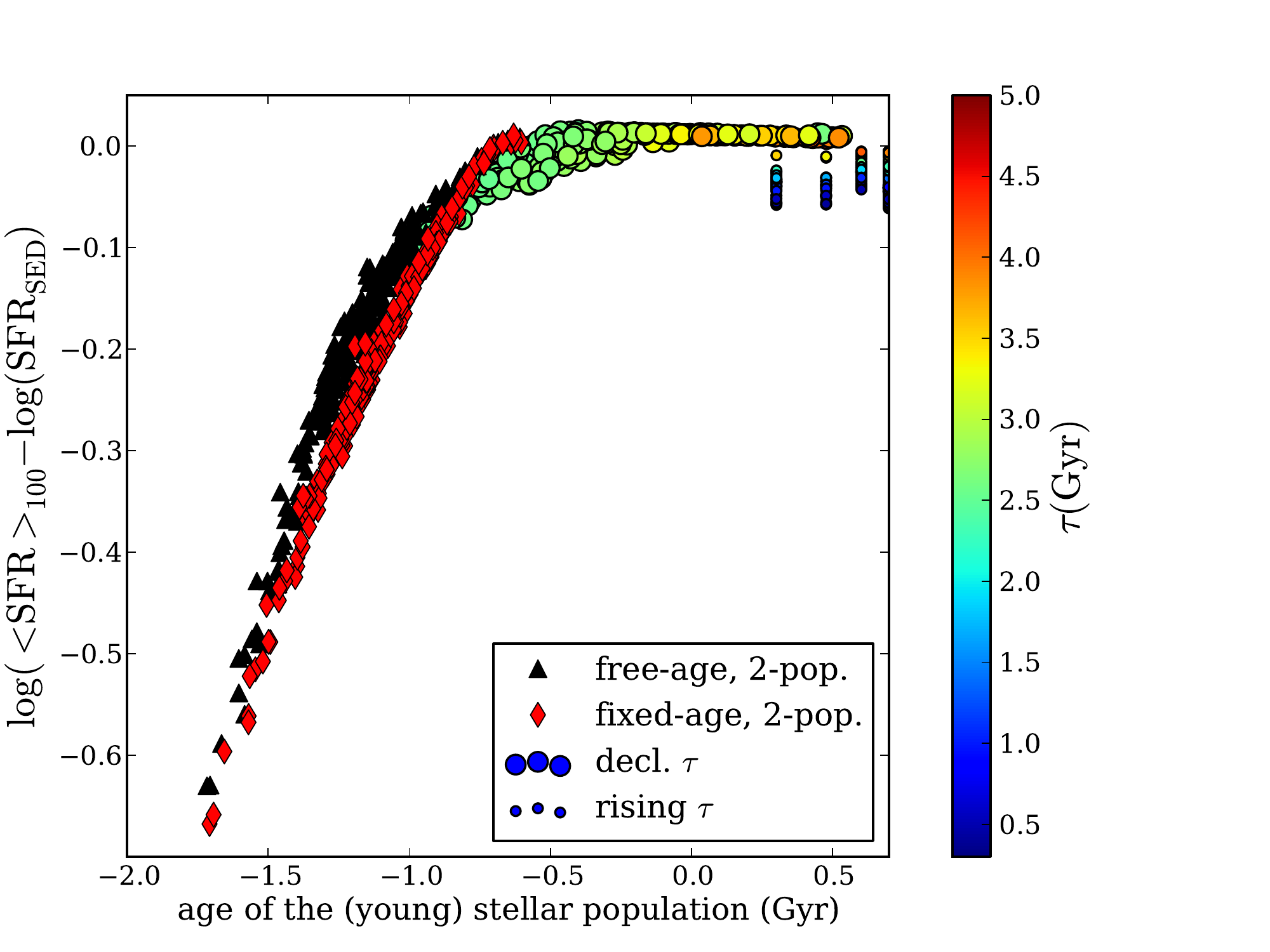}
   \caption{Difference between the SFR averaged over 100 Myr and the instantaneous one given by CIGALE plotted as a function of the age of the youngest stellar population ($t_f$ for the $\tau$-models and $t_{\rm ySP}$  for the two-populations model). Triangles are for the two-populations model, small circles  for the  rising-$\tau$ model and large  circles for the $\tau$ model. The absolute value of the decreasing rate $\tau$ is color coded  for $\tau$-models. }
              \label{sfr100}%
    \end{figure}

 \subsubsection{Star formation rate calibrations}
We now   compare the SFR found with our fitting method ($SFR_{\rm SED}$) to the one deduced directly from  recipes converting  FUV and IR luminosities  into SFRs. The standard calibrations assume  a constant SFR over $\sim 10^8$ years, which is sufficient to  reach a stationary state for the UV production.  Under this assumption, the intrinsic FUV luminosity is directly proportional to the current SFR.   The total IR ($\rm \sim 5 - 1000 \mu m$) luminosity is related to the SFR  by assuming that the bolometric emission of  young  stars is absorbed by dust and re-emitted in IR (\cite{kennicutt98}). Here we consider the total SFR as the sum of SFR  obtained with the IR and FUV (not corrected for dust attenuation) luminosities.  We do not account for  heating of dust by older stars since all  stellar populations are  quite young.  \cite{buat08}  obtained a calibration   for a Kroupa IMF  
  and a constant SFR over $10^8$ years. The observed FUV luminosity is  calculated at 1530 $\AA$ and the total IR luminosity is  integrated over the whole IR SED. It is an output of the  CIGALE code,  which is  found independent on the assumed SFH. The SFR is calculated with the formula:
  $$SFR_{\rm IR,FUV} = SFR_{\rm IR}+SFR_{\rm FUV} = L_{\rm IR}/10^{9.97}+L_{\rm FUV}/10^{9.69}$$ 
  where the SFR is expressed in $\rm M_\odot yr^{-1}$  and the luminosities in $\rm L_\odot$
  \\
 The comparison   between  $SFR_{\rm IR, FUV} $ and  $SFR_{\rm SED}$ is shown  in Fig.\ref{comp-SFR}. Both estimates are very consistent for $\tau$-models since  most of the sample stellar populations are  older than   $10^8$ years with a SFH close to a constant one.  There is a  small shift between $ SFR_{\rm IR,FUV}$ and  $SFR_{\rm SED}$  for the two-populations model:   the young population is  younger than $10^8$ years ($t_{\rm ySP} = 76 \pm 38$ Myr)) and the UV production does not reach a fully steady production rate, leading to a slight underestimate  of the instantaneous SFR with $ SFR_{\rm IR,FUV}$. The difference is tightly correlated to $t_{\rm ySP}$, the age of the young stellar population, but  remains very modest  on average  ($0.05\pm 0.03$ dex)  (Fig.\ref{comp-SFR}, lower panel). The same effect is observed for decl.-$\tau$ models with very young stellar populations, leading to $SFR_{\rm SED}$ larger than $ SFR_{\rm IR,FUV}$  (Fig. \ref{comp-SFR}, upper panel).
 The few galaxies for which $SFR_{\rm SED}<SFR_{\rm IR,FUV}$ correspond to objects with a very low mass fraction locked in the young stellar component ($\simeq 2\%$): the SFH is dominated by the decreasing old component, which  is not  constant  over the last $10^8$ years  but,  instead,  has  slightly decreased.  \\
 CIGALE  also provides the  SFR averaged over the last 100 Myr $<SFR_{100}>$.  The difference between $<SFR_{100}>$ and $SFR_{\rm SED}$   are  plotted   in Fig.\ref{sfr100} against the age of the (youngest) stellar population.  For  $\tau$-models,  the difference between the averaged and instantaneous  SFR  depends on the age of the  stellar population and on the absolute value of e-folding rate $\tau$, but the mean difference remains very  small (-0.01 dex for  the decl.-$\tau$ model, -0.03 dex for the rising-$\tau$ one). The situation is very different for the two-populations model:  the difference between the averaged and instantaneous  SFR strongly depends on the age of the young stellar population, which dominates the current star formation. The mean difference reaches -0.20 dex, but much larger differences can be found for  individual objects. Very similar trends are found when $ SFR_{\rm IR,FUV} $ is used  instead of the instantaneous SFR (not shown here).  This  clearly demonstrates that the SFR measured  with UV and IR data, either with  SED fitting or with empirical calibrations,   are not     equivalent to a SFR averaged over 100 Myr.

\subsection{Star formation rate-stellar mass relation}

\begin{figure*}
\centering
\includegraphics[width=15cm]{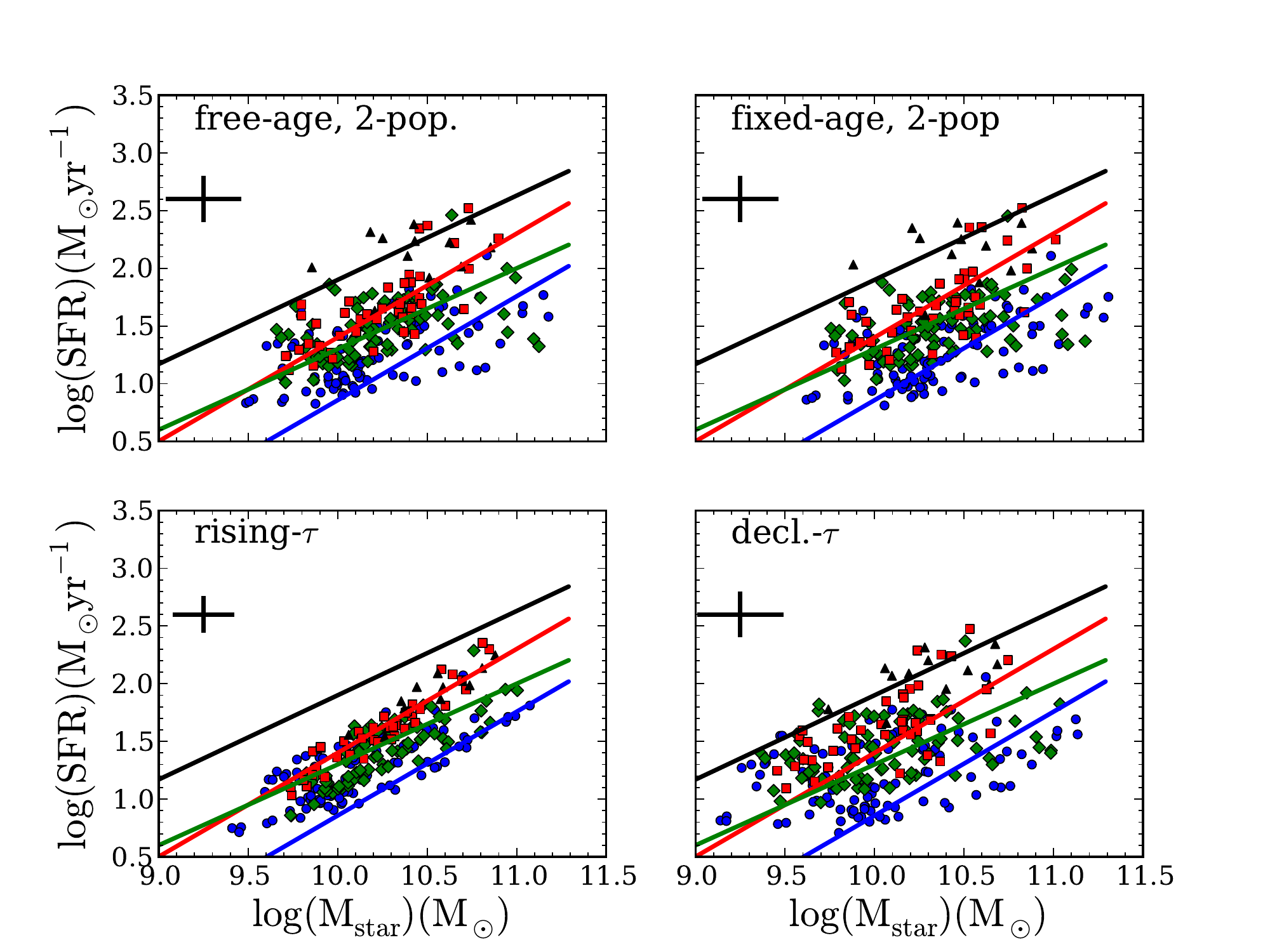}
\caption{ SFR versus  $M_{\rm star}$ (left panels).  Blue points $z<1.2$, green diamonds $1.2<z<1.7$, red squares $1.7<z<2$, black triangles $z>2$. From top to bottom: two stellar populations, one decreasing population, one increasing population models. The relations proposed by   \cite{elbaz07} (blue, $z\sim 1$), \cite{daddi07} (red, $z\sim 2$) and \cite{heinis13b} (green, $z=1.5$, black $z=3$) are over  plotted with solid lines.}
\label{MS-allSFH}%
 \end{figure*}

We have shown that the  values of  SFR and $M_{\rm star}$ depend on the assumptions made to derive them. We now explore the consequences of these variations on the SFR-$M_{\rm star}$ scatter plot.  The SFR-$M_{\rm star}$ relation is expected to evolve with redshift (\cite{noeske07,wuyts11,karim11}).
 We  split the sample into four  redshift bins $z<1.2$, $1.2<z<1.7$, $1.7<z<2$, and $2<z<3$.  The SFR and $M_{\rm star}$ deduced from the SED fitting are plotted  in Fig. \ref{MS-allSFH}. Very similar plots are found using $SFR_{\rm IR, FUV} $  as expected from the very good correlation found between both SFR estimates. Different relations from the literature and  covering our redshift range are overplotted: the relations of \cite{elbaz07}  and \cite{daddi07} at $z=1$ and 2  respectively, and the relations of \cite{heinis13b} at $z=1.5$ and 3. 

 Slight  variations   are  seen  between both two-populations and decl.-$\tau$ models, caused by  the difference in $M_{\rm star}$ measurements.  The dispersion is large in all the scatter plots (except for the rising-$\tau$ model as discussed below) and we do not try to provide  any  relation.  Our galaxy sample is built to have the best wavelength coverage, but is not complete in any sense and is not suited to derive a representative SFR- $M_{\rm star}$ relation.  Our selection of galaxies observed both in FUV and IR (rest-frame) is likely to bias towards objects with a  high star formation activity. 
Free-age and fixed-aged two-populations models lead to a similar dispersion with a slight shift towards larger $M_{\rm star}$ and lower specific star formation rate (SSFR) for the fixed-age model. 
 Lower $M_{\rm star}$   and similar SFR  are obtained with the decl.-$\tau$-model  as compared to the results obtained with   the other models and the discrepancy is larger when $M_{\rm star}$ decreases (Section 3 and Fig.\ref{Mstars}).   
 It induces  a flatter  SFR-$M_{\rm star}$ relation  as compared to the other models. The residual RMS error obtained with a linear fit for each redshift bin is $\sim 0.3$ for both two-populations and decl.$\tau$ models.  The rising-$\tau$ model leads  to a steeper, well-defined relation  SFR-$M_{\rm star}$, which does not much evolve with redshift and the residual RMS error to linear fits is reduced to $\sim 0.2$. This reduction  is expected since  rising SFHs have been introduced to explain the small scatter found in the SFR-$M_{\rm star}$ relation and the  non-evolution of the SSFR at very high redshift (\cite{maraston10,papovich11}). This variation of the RMS error     illustrates that the tightness of the derived SFR-$M_{\rm star}$ relation also depends on the way these two quantities are derived. 
  
 \section{Fitting spectral energy distributions with a reduced dataset}
 
\begin{figure*}
\centering
\includegraphics[width=17.5cm, height=12.5cm]{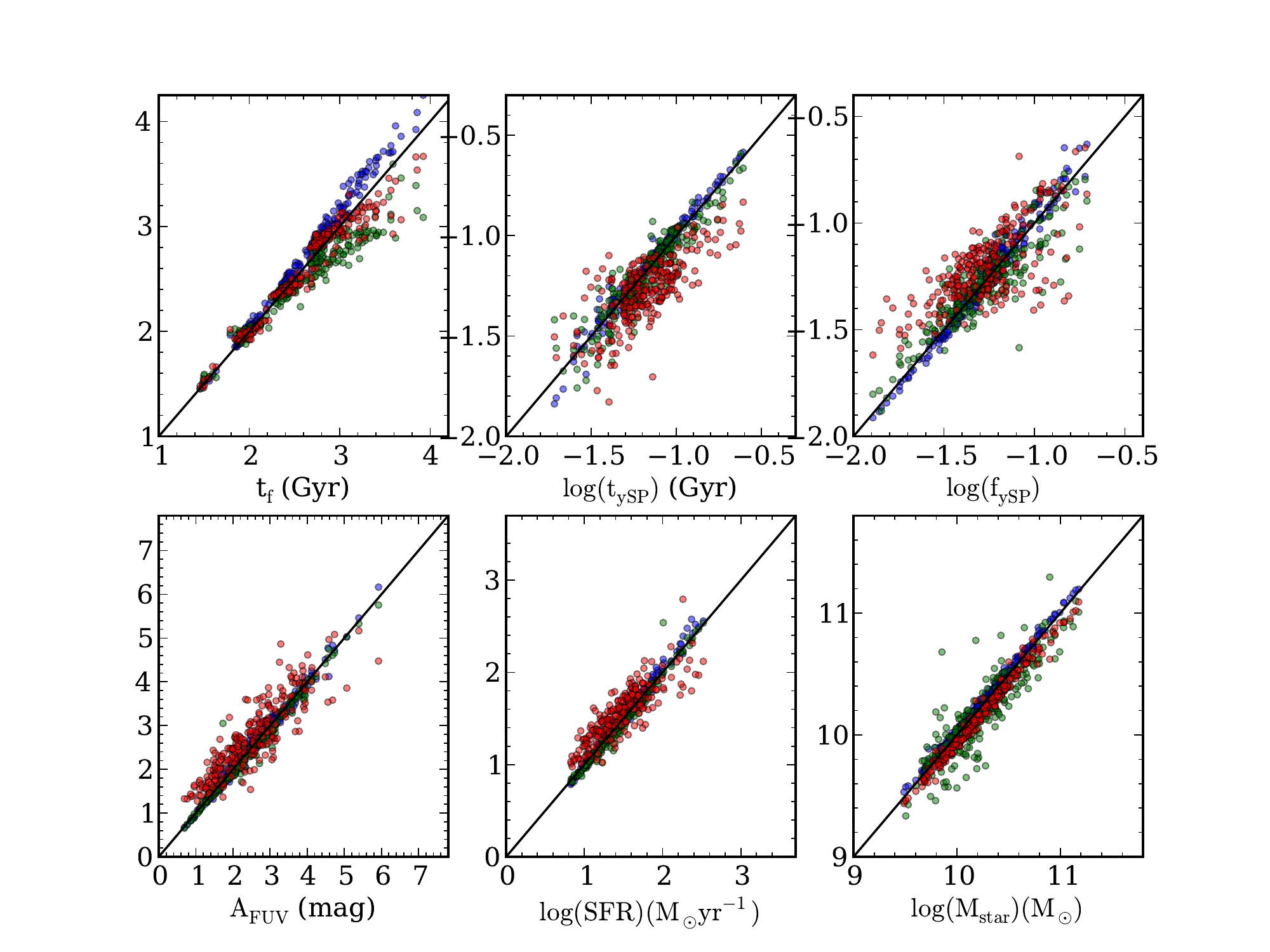}
\caption{ Comparison between parameter estimations  for  the  two-populations model  with and without  IB, NIR, and IR data. X-axis: all the dataset is used.  Y-axis: parameters estimated  without IB data (blue circles),   without NIR  data (green circles), and without IR data (red circles), }
\label{comp-bands}
\end{figure*}
  
We now check the importance of the wavelength coverage to estimate our physical parameters. We identify three sets of data whose influence has to be investigated: the IB filters, sampling the UV continuum and sensitive to the recent SFH; the NIR rest frame data (K and all IRAC bands), often presented as the best tracers of $M_{\rm star}$; and the IR emission from dust  sampled by IRAC4, MIPS, and PACS data, which is expected to provide a strong constraint to dust attenuation. We perform  fits (i) omitting data obtained with all  IB filters, (ii) omitting NIR data  and (iii) omitting IR data. When one dataset is omitted, all the other data are kept. It may appear unrealistic from an observational point of view since galaxies detected in the IR are all detected with IRAC. Nevertheless, our aim is to test the influence of a specific wavelength domain on the determination of the physical parameters and to disentangle the different potential biases.
The tests are  performed for all SFH models, but are  presented here for the two-populations model only. Similar  trends are found for the  other models.

\begin{figure}
   \centering
\includegraphics[width=\columnwidth]{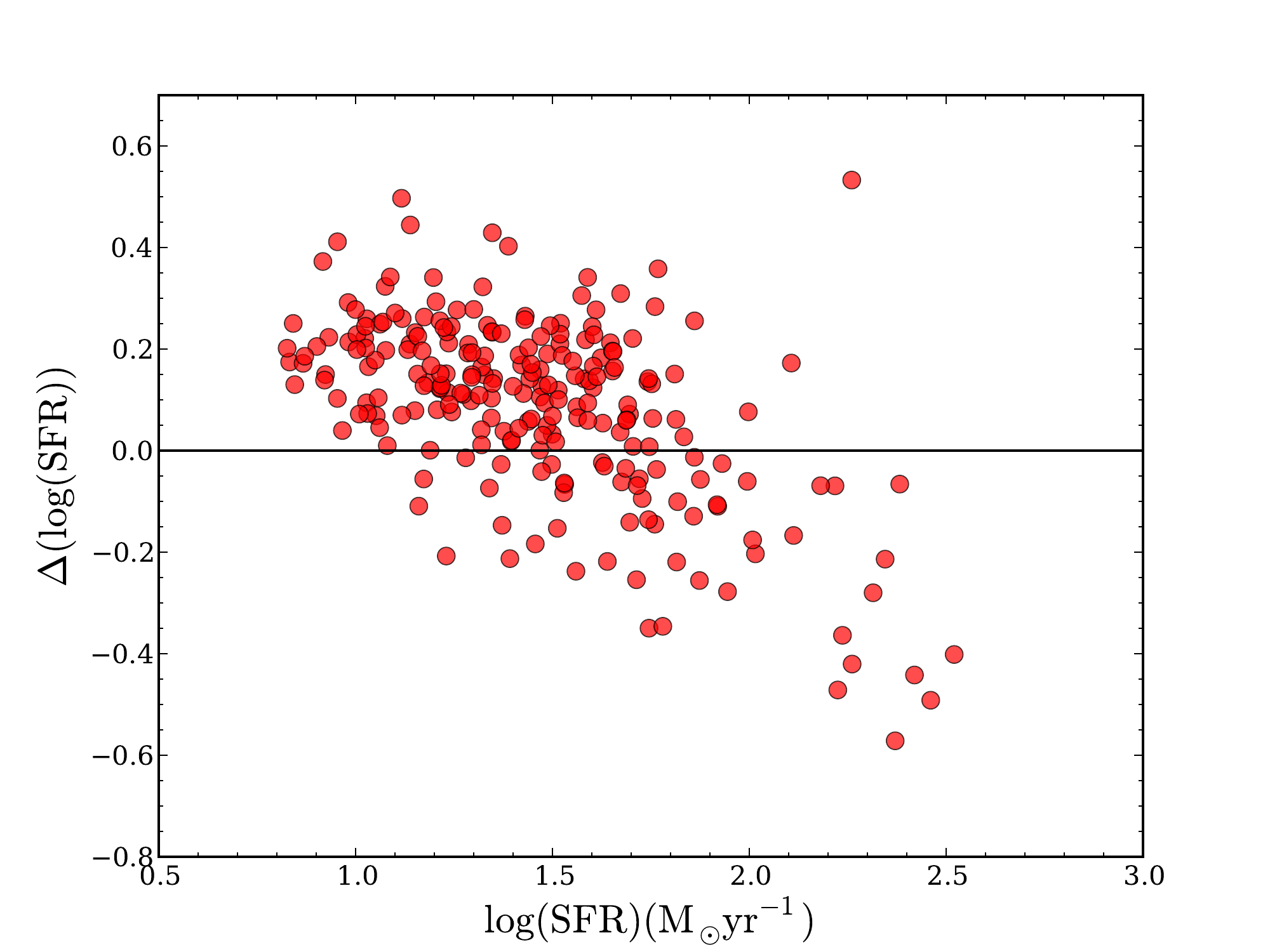}
\includegraphics[width=\columnwidth]{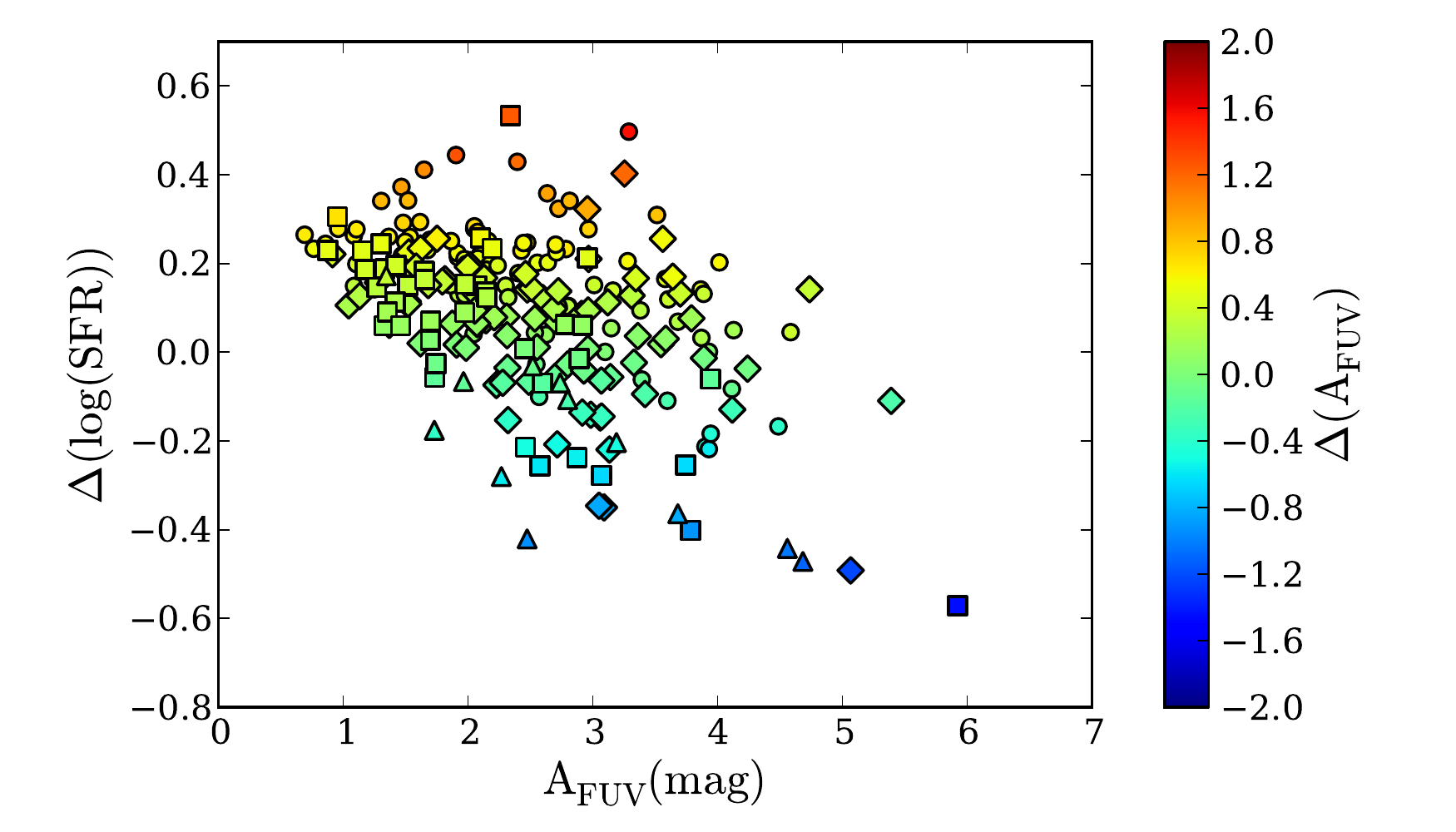}
 \caption{ Difference in SFR estimates when IR data are omitted. Upper panel: The SFR reported on the x-axis is calculated with the whole dataset.  The $\Delta(\log(SFR)$ on the y-axis  is defined as the $\log(SFR)$(with the reduced dataset)-$\log(SFR)$ (with the whole dataset). Lower panel: $\Delta(\log(SFR)$  as a function of $A_{\rm FUV}$ measured with the full dataset. The difference between $A_{\rm FUV}$ is color coded. }
  \label{deltasfr}%
 \end{figure}

In Fig. \ref{comp-bands},  the estimated values  of the age of the stellar populations ($t_{\rm f}$ and $t_{\rm ySP}$), the mass fraction in the youngest population ($f_{\rm ySP}$), the dust attenuation in FUV ($A_{\rm FUV}$), SFR and $M_{\rm star}$, obtained  for cases (i), (ii), and (iii), are compared to those obtained with the full dataset (from  Section 3).  With CIGALE, the internal accuracy of the different parameter estimations  is  measured   by the   standard deviation  of  the probability distribution function of each parameter and for each object.  The mean  values of this standard deviation,  averaged  over  the full  sample of 236 galaxies are listed in Table 2 for  each dataset. \\
All  parameter estimates and the associated  averaged dispersions  are  very similar with and without IB data, which  play a minor role in these estimations.  Only  the values of $t_{\rm f}$ are  slightly larger without IB data, with no  consequence in the determination of SFR and $M_{\rm star}$. We recall that the IB filters  cover only the UV rest-frame range, which is essentially featureless. The situation is likely to be very different for IB sampling of the visible range with  strong emission lines, which can strongly modify the overall spectrum (\cite{schaerer13}). A good sampling of the UV continuum is useful  to constrain the dust attenuation curve (Paper I) but does not add any useful information on the recent SFH provided that broadband photometry is available to measure the UV emission. \\
The decision of whether or not to include the NIR rest-frame data modifies most of the parameters, except $A_{\rm FUV}$ and SFR, which are exclusively related to the young stellar population. We find moderate  shifts  for the ages of the stellar populations. When NIR data are excluded, $t_{\rm f}$  is lower by up to 0.5 Gyr  when $t_{\rm f} > 2.5$ Gyr and the young stellar populations are found to be 10\% younger. The impact on  the determination of $M_{\rm star}$  is  a small systematic shift,  $M_{\rm star}$ being lower by $15\%$  on average when NIR data are excluded,   with a dispersion reaching 0.16 dex between estimations with and without NIR data.  The intrinsic uncertainty in the determination of $M_{\rm star}$ also increases from 0.21 dex for the full dataset to 0.33 dex without NIR data.\\
 Without IR data, results substantially change for all parameters linked to recent star formation. As expected, the parameters  related to the old component  ($t_{\rm f}$, $M_{\rm star}$) are much less affected. A substantial dispersion is observed in Fig. \ref{comp-bands} between the values  of $t_{\rm ySP}$,  $f_{\rm ySP}$,  SFR and $A_{\rm FUV}$, estimated with and without IR data (corresponding to 0.14 dex, 0.16 dex, 0.18 dex, and 0.44 mag, respectively).   Looking at Table 2 it is clear that omitting IR data  affects the accuracy  of   $A_{\rm FUV}$ and  SFR.    The distribution of all these parameters  is  flatter without IR data and SFR are larger by  $20\%$  on  average when IR data are excluded. This  is illustrated in Fig.\ref{deltasfr} with a dependence  of the SFR estimates when IR data are not present   on the  SFR   measured  with the whole dataset. Under the assumption  that SFR  obtained  with the whole dataset are reliable, in the absence of  IR data,   low SFR  are overestimated  and large SFR  are underestimated,  by a factor that can reach 2.5. The difference in SFR estimations strongly depends on the estimation of dust attenuation  as seen in the lower panel of  Fig.\ref{deltasfr}. If we assume that secure measurements are obtained when including IR data, large attenuations are  underestimated  implying  an underestimation of the SFR, the inverse trend being observed for low attenuations. 
 This  is  consistent with  the conclusion of  \cite[and erratum]{burgarella05} that without IR data, the SED fitting process biases towards average values of dust attenuation. 

\begin{table}
 \caption{Intrinsic uncertainty of the parameter estimation using the different datasets, measured as the dispersion of the probability function of each parameter. }
\label{tab:dispersion-parameters}
\centering
\begin{tabular}{c c c c c }
\hline
{\bf parameter }&full data & no IB & no NIR & no IR\\
& & (i)&(ii)&(iii)\\
\hline
$t_f$(Gyr) & 1.09 &1.07&1.10&1.10\\
$\log(t_{\rm ySP})$ (Gyr) & 0.45 & 0.44 & 0.47& 0.49\\
$\log(f_{\rm ySP})$ & 0.49 & 0.48 & 0.52 & 0.53 \\
$A_{\rm FUV}$ (mag) & 0.52 & 0.47 & 0.54& 0.86 \\
$\log(SFR) ({\rm M_\odot yr^{-1}})$ & 0.18 & 0.16 & 0.19 & 0.33 \\
$\log(M_{\rm star}) ({\rm M_\odot yr^{-1}})$ & 0.21 & 0.19 & 0.33 & 0.24 \\
\hline
\end{tabular}
\end{table}

 \section{Mock galaxies}
We checked the internal consistency and limitations of the method we  used to retrieve physical quantities of galaxies. To this aim we built a sample of  artificial SEDs defined in  the same filters as those used for  real sources. Then we attempted to retrieve the known properties of these pseudo-galaxies and to check the reliability of parameter estimation as well as the presence of potential systematic biases. 

\subsection{Definition of the mock catalogue with  $1<z<2$ and fitting process}
Given the few sources with $z>2$, we restrict the mock catalogue to $1<z<2$.  The actual  redshift is not  important since the models are only dependent on  the physical ages of the stellar populations. The redshift range is  introduced to reproduce data corresponding to the real sample and to account for the redshifting of the rest-frame emission of galaxies. \\
We only consider   the free-age two-populations model. If we were to  use all the parameters considered in Table 1  to simulate the artificial SEDs the total number of sources is very  large and the computational time prohibitively long. Therefore, we restrict the sampling of the input parameters to the values appearing with bold characters in Table 1.  We choose a  single  distribution of dust emission  ($\alpha=2$), the dust attenuation law is fixed with parameters chosen from our previous study (Paper I). The sampling of the amount of dust attenuation is also  reduced (but with the same total amplitude of variation), and we limit the mass fraction in the young stellar population  to 20$\%$ at most. The SEDs are generated for all of the 30 bands (including the PACS bands) considered in this work, and a random noise, typically of the order of 10$\%$,  is added to the simulated fluxes.  A similar, relatively small, error is applied to all the data since our aim is not to measure the effect of varying observational conditions,  but to understand the limitations intrinsic to our modelling. All galaxies are created with  a total galaxy mass of $10^{10} M_\odot$:  we  cannot check the dynamical range  of SFR and  $M_{\rm star}$ but only  systematic changes for these parameters. A total of 4970 artificial galaxies are created.\\

We fit the mock data, using CIGALE with the whole  set of input parameters, as for real galaxies and not using  the reduced set used to create the catalogue (cf. Table 1)
We again perform again  four distinct fits: using the whole dataset,  excluding IB, NIR,  or IR data. Note that when considering the influence  of NIR rest-frame  data we are not testing the role of the stellar population models since we are creating and fitting our artificial data with the same code.\\

\begin{figure}
   \centering
   \includegraphics[width=\columnwidth]{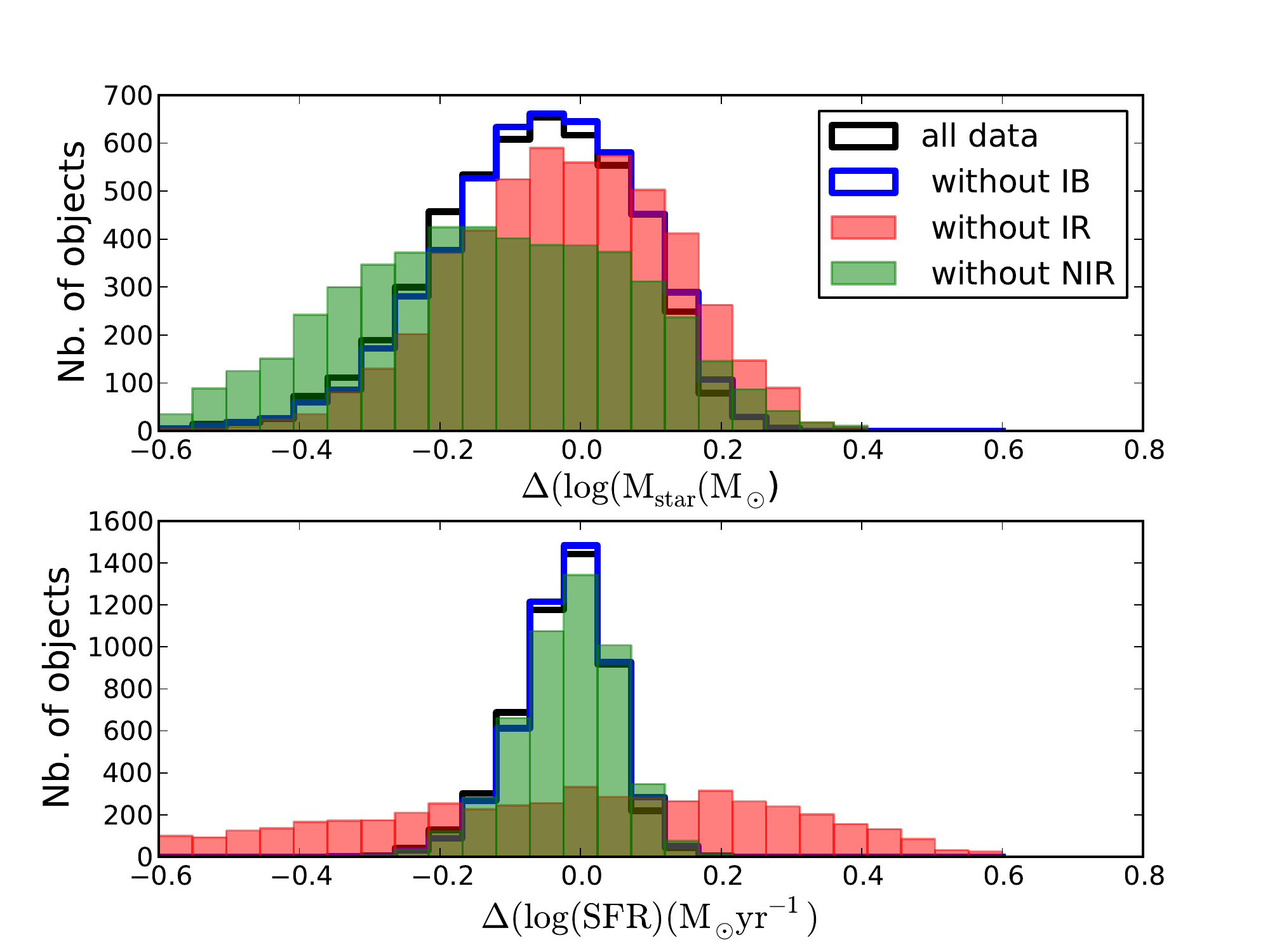}
\includegraphics[width=\columnwidth]{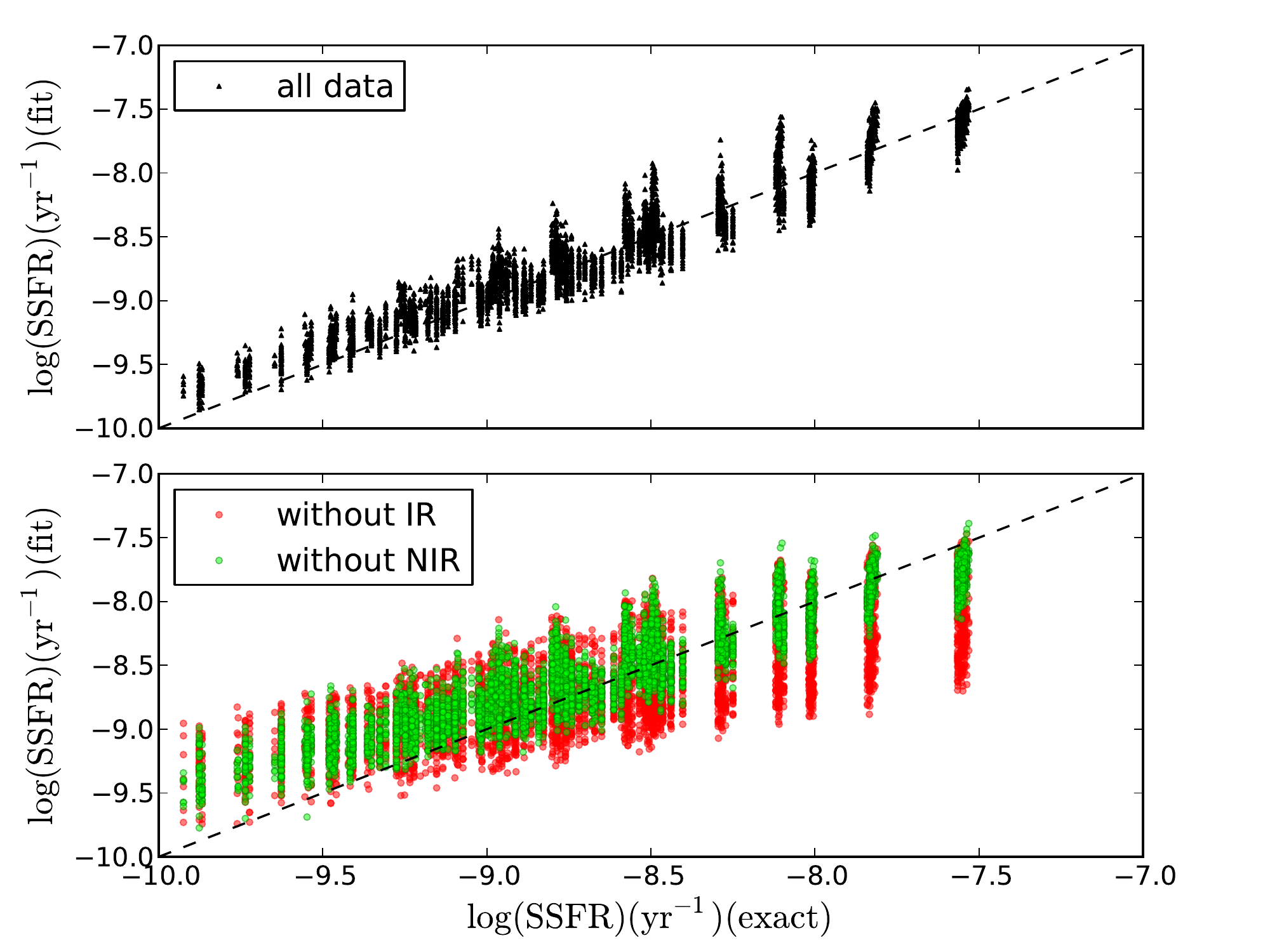}
     \caption{Upper panels: Histogram of the difference between the estimated values of $M_{\rm star}$ and SFR  and the exact values from  the mock catalogue. Black line: whole dataset, blue line: without IB data, green filled histogram: without NIR data, red filled histogram: without IR data,  Lower panel:  SSFR estimated with SED fitting are compared to the true values of the SSFR. Black point: whole dataset,  green dots:  without NIR data, red dots: without IR data. Results without IB are not shown since they are indistinguishable from   those without IR data. }
              \label{Mstarhist}%
    \end{figure}

\subsection{Stellar mass and star formation rate determinations. }

In Fig. \ref{Mstarhist} the difference between all the estimated and true values of SFR and  $M_{\rm star}$  are represented for the three combinations of data\footnote{In Section 5, all the difference between the estimated and true values of a parameter   are defined as 'estimated value-true value'.}. No effect on $M_{\rm star}$ and SFR is found due to the lack of introduction  of IB data, which confirms the findings  with  the real sample, the case without IB data will not be discussed. The impact of IR data is clearly seen, especially for the SFR determinations and the lack of introduction  of NIR data modifies the  $M_{\rm star}$ distribution but in a much smaller amount than the IR data for the SFR.  \\
When all the data are considered, there is only a small   systematic difference between the estimated and true values of $M_{\rm star}$: $<\Delta(\log(M_{\rm star})> =-0.07$ dex with a   $1\sigma$ dispersion  equal to  0.14 dex. Without NIR data  the systematic shift and the dispersion are larger: $<\Delta(\log(M_{\rm star})> =-0.13$ dex and $ \sigma=0.20$ dex. We confirm the difference of  15$\%$ found in Section 4.1,  with and without NIR data for real galaxies.    \\
The situation is worse for the SFR, when IR data are omitted: if the mean systematic difference, $<\Delta(\log(SFR)> $,   remains consistent with 0, the dispersion varies from 0.05  dex with IR data to 0.30 dex without IR data. This poor  determination of the SFR without IR data is clearly visible when SSFR  are compared  (lower panels of Fig. \ref{Mstarhist} ). We confirm that the range of values of SSFR  found without IR data is reduced as compared to that of the true values, which are well estimated when IR data are considered. Intrinsically,  small SSFR  are overestimated without NIR data, this is due to  underestimation of stellar masses for these objects, since SFR  are well estimated in this case. 
    \begin{figure}
   \centering
  \includegraphics[width=\columnwidth]{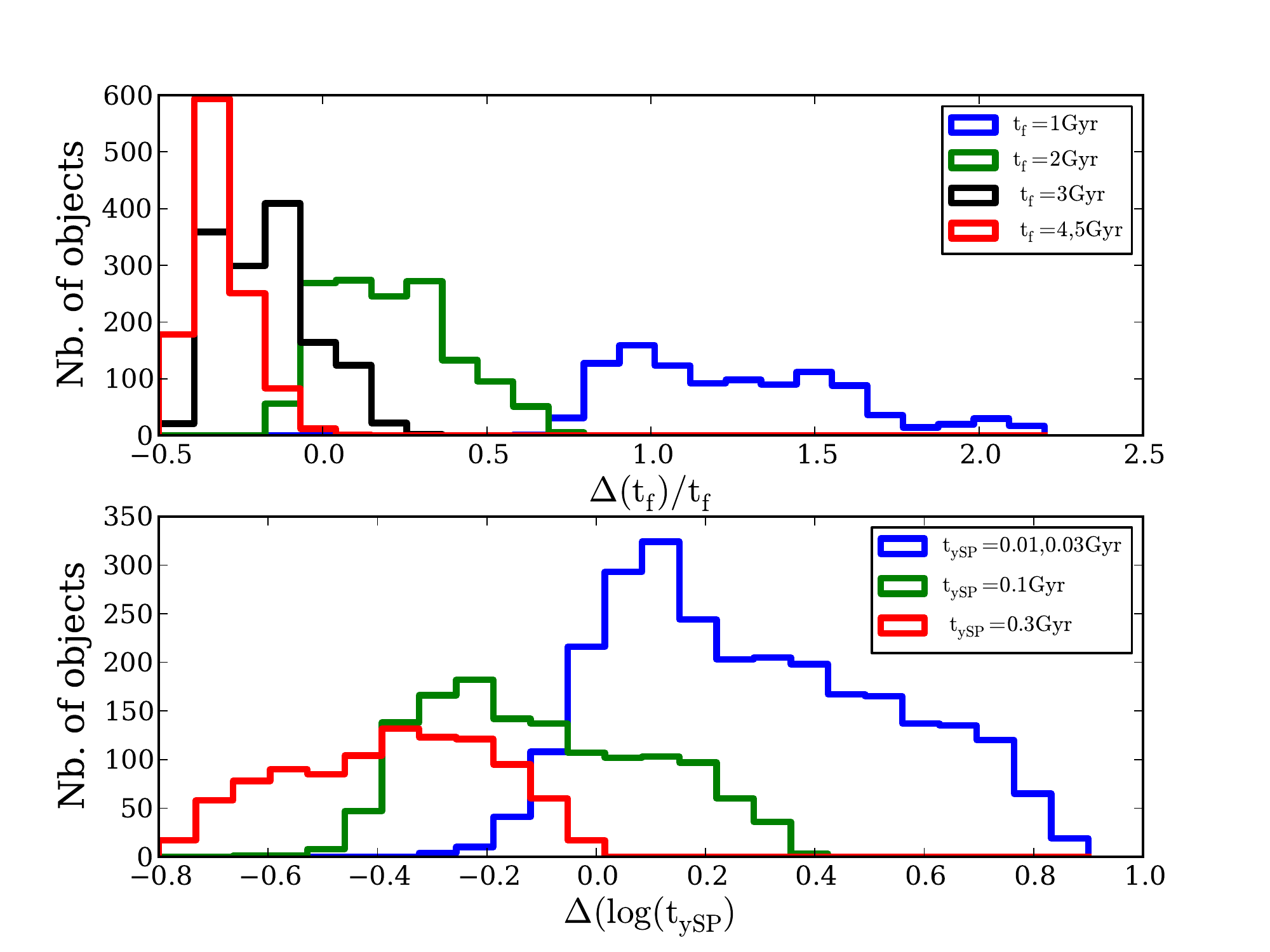}
   \caption{Upper panel: distribution of the relative difference between the estimated and true  age of the old stellar population. The sample is split as a function of the true age: $t_f= 1$ Gyr, $t_f=2$ Gyr, $t_f=3$ Gyr,  and $t_f=4,5$ Gyr.  Lower panel: distribution of the ratio of the estimated and true age of the young stellar population in logarithmic units. The sample is split as a function of the true age: $t_{\rm ySP} = 0.01$ Gyr, $t_{\rm ySP}= 0.03$ Gyr , $t_{\rm ySP}=0.1$ Gyr,  and $t_{\rm ySP}=0.3$ Gyr.  }
              \label{age-mock}%
                 \end{figure}
 \begin{figure}
   \centering
  \includegraphics[width=\columnwidth]{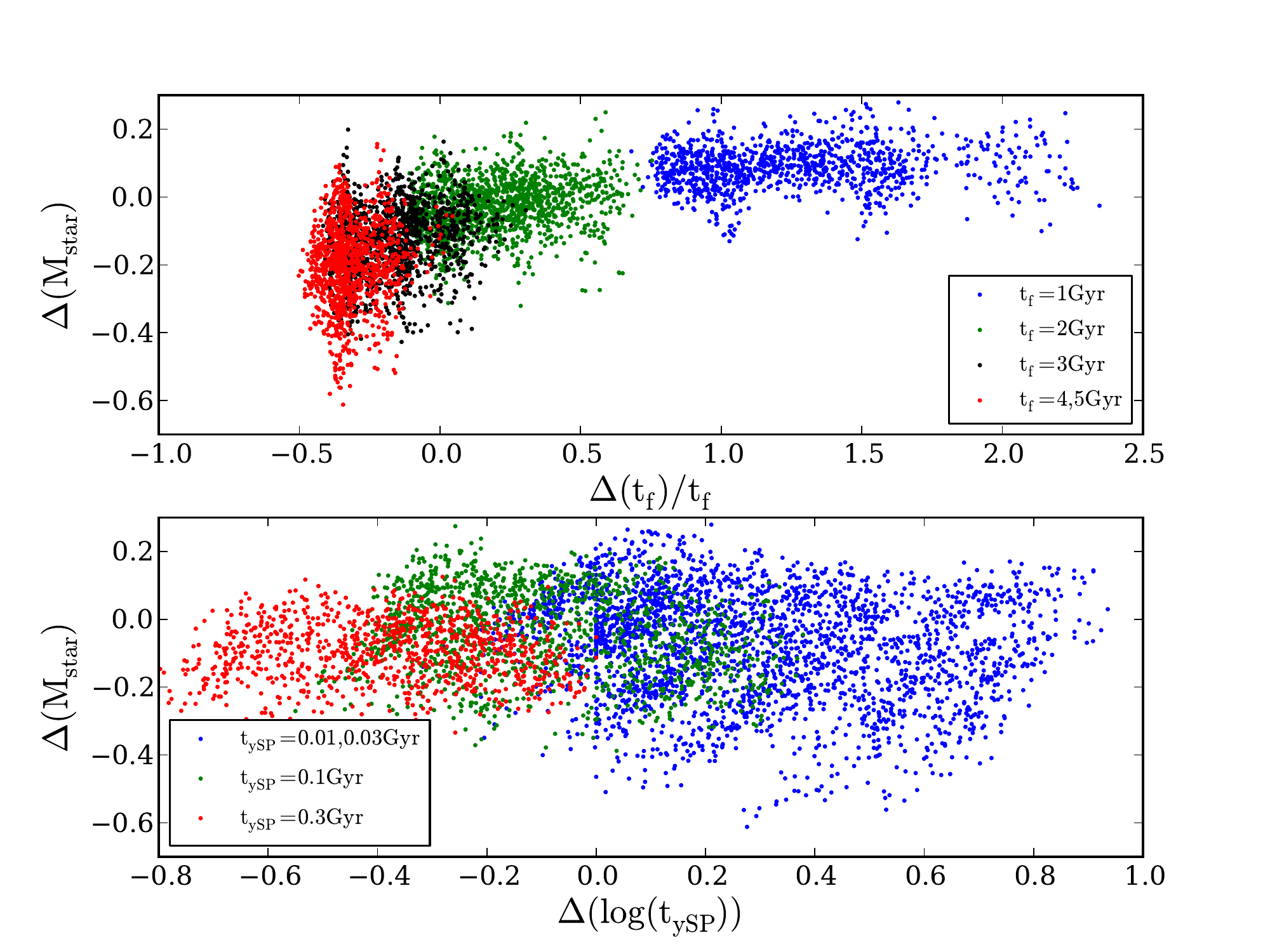}
    \caption{Difference between estimated and true values of SFR and $M_{\rm star}$ as a function of the difference between estimated and true ages. The colors are the same as in Fig.\ref{age-mock}}
    \label{mock-2}
    \end{figure}
    \begin{figure}
   \centering
   \includegraphics[width=\columnwidth]{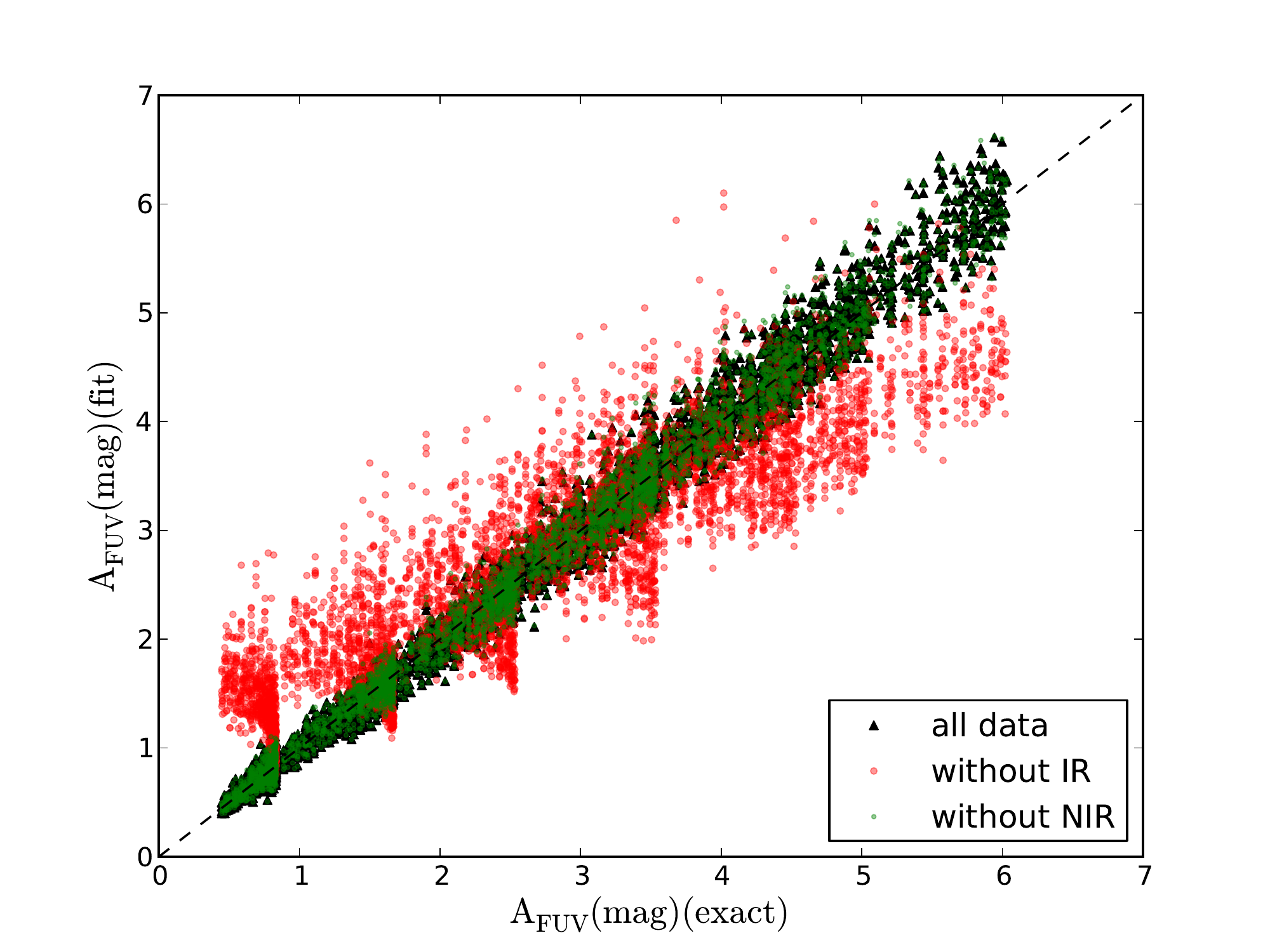}
  \includegraphics[width=\columnwidth]{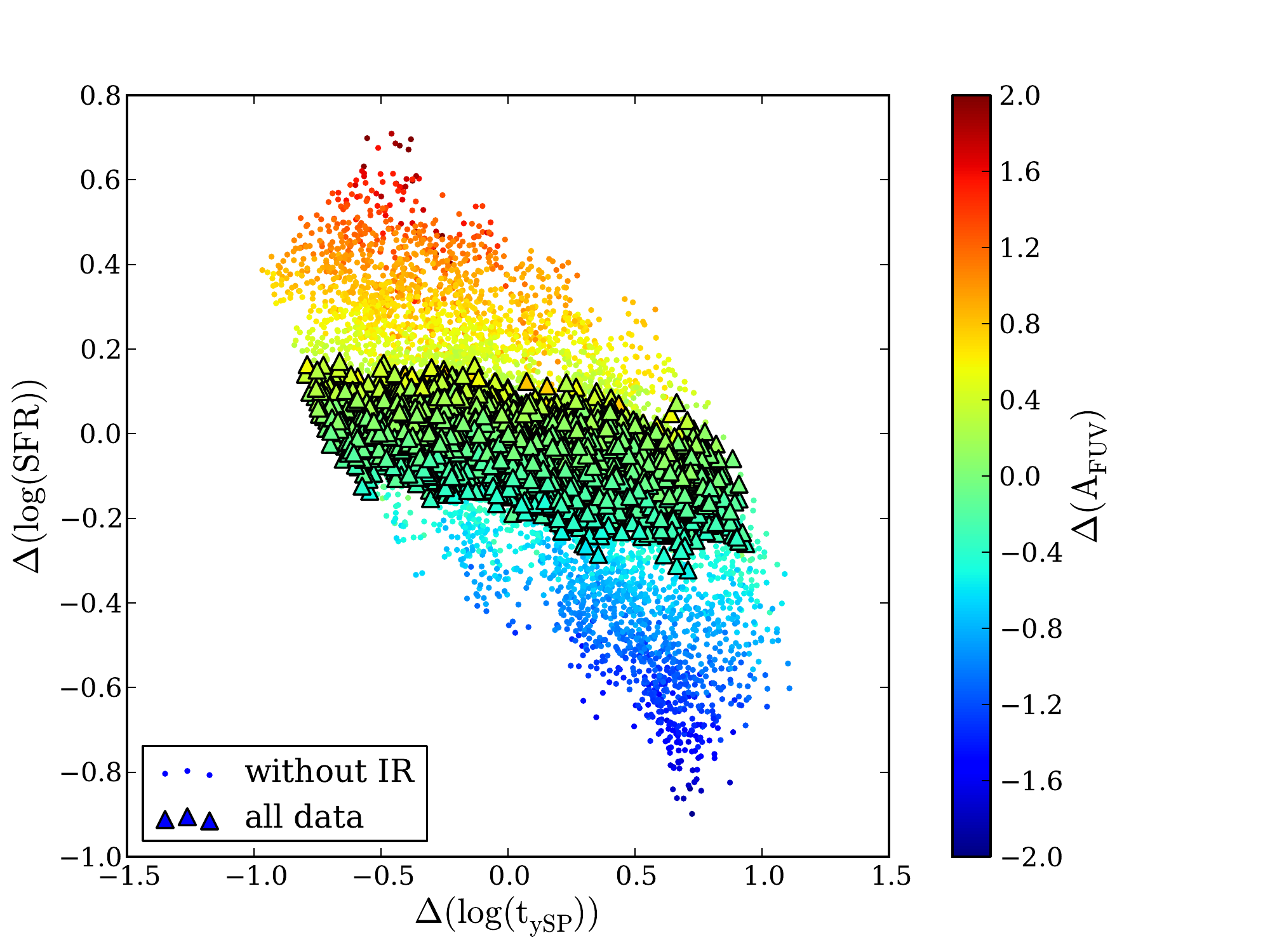}
 \caption{Difference between estimated and true values of SFR  as a function of the difference between estimated and true ages, dot are for fits with IR data, crosses correspond to fits without IR data.  The difference  between estimated and true values of $A_{\rm FUV}$ is shown with a color scale.}
    \label{age-dust-sfr-mock}
    \end{figure}
    
\subsection{Stellar ages}
We explored the different estimations of  stellar ages,  with and without IR and NIR data. We found very similar trends whether IR data are omitted or not. Therefore, in this sub-section, we  begin by studying the fit with the whole dataset, and then we focus on  the difference found between the fits with and without NIR data.\\
We have seen that the  determination of the age of the stellar population is  crucial to describing the SFH and to measuring  $M_{\rm star}$, and that the impact on SFR is less important, as discussed in  Section 3 and in the next sub-section.  Our artificial sources are created with  stellar populations spanning  a large range of ages,  between 1 and 5 Gyr for the old component and 0.01 to 0.3 Gyr for the young component.
The relative difference between the estimated and true values of $t_{\rm f}$  is shown in Fig.\ref{age-mock}. Ages lower than $\simeq 2$ Gyr are overestimated (by 127$\%$ for $t_{\rm f}=1$ Gyr  and by 20$\%$ for $t_{\rm f}=2$ Gyr)  whereas larger ages are underestimated (by 16$\%$ for $t_{\rm f}=3$ Gyr  and by 30$\%$ for $t_{\rm f} =$ 4 and 5 Gyr). The strongest discrepancy is found for the very small  ages of  the oldest stellar population.  \cite{lee09} also found that  the ages of model Lyman Break Galaxies , at $z>3$  and lower than 1 Gyr,  were strongly overestimated with SED fitting methods. We  confirm that  the ages of the old stellar populations are underestimated in typical  galaxies with an old ($> 3$ Gyr) underlying stellar population. The same effect is found for the age of the young stellar population.  \\
These systematic shifts in age determination have a direct impact on  $M_{\rm star}$ measurements as shown in Fig.\ref{mock-2}.  Underestimations of $t_{\rm f}$ and $M_{\rm star}$ are correlated,  which explains  the large dispersion found in the $M_{\rm star}$ distribution. The effect of $t_{\rm ySP}$ is less important, as expected, but the presence  of  a very young stellar population increases the uncertainty in the mass determination (Fig.\ref{mock-2}, lower panel).\\

\subsection{Dust attenuation}
We have seen in Section 4 that dust attenuation has a major impact on SFR determinations. We re-investigate its role with our mock catalogue. 
The estimates of  $A_{\rm FUV}$,  with and without IR data, are compared in Fig.\ref{age-dust-sfr-mock}.  Whereas  dust attenuation is very well measured with IR data, when these IR data are missing  $A_{\rm FUV}$ is overestimated in systems with low attenuation and underestimated in very dusty systems, confirming the flattening of the  distribution of $A_{\rm FUV}$  found in Section 4 in the absence of  observed IR data.  It can be seen in Fig.\ref{age-dust-sfr-mock} that dust attenuation  plays a crucial role for SFR determinations whereas the age of the  stellar population dominating the current SFR ($t_{\rm ySP}$)   only acts as a secondary parameter.
 When this age is overestimated the recent star formation is diluted over too large atimescale and the SFR is underestimated, as also found by \cite{lee09}, but the effect remains very modest for our sample. 
\\

\section{Conclusions}
We  measured SFR, $M_{\rm star}$, and stellar ages for  a  sample  of 236 galaxies at $1<z<3$   in the GOODS-S field observed with an excellent  wavelength coverage, including UV and IR rest-frame data. We used 28 photometric bands   from U to 24 $\mu$m, and 80 sources are detected with PACS at 100 or 160 $\mu$m. We also considered  intermediate band (IB)  filters which sample the UV rest-frame. We performed SED fitting with the code CIGALE, which implements an energy budget between  dust and stellar emission. We   explored different SFH: exponentially increasing and decreasing $\tau$-models and a model combining an old decreasing SFR and a younger population of constant SFR. In a first step, the age of the main stellar population was left  either fixed or free. The Decl.-$\tau$ models with a fixed redshift formation ($z_{\rm f}\simeq 8$) did not fit the data well. All the other  models yield much better results at the price of young ages for free-age models, which were   unrealistic with $\tau$-models.  The best fits are obtained with two-populations models. Ages were   higher for  the free-age two-populations model.  In a second step,  the models selected for the study  were  the free-age decl.$\tau$ model (because of its popularity), the fixed-age rising-$\tau$ model, and both fixed and free-age two-populations models. We investigated the impact of the coverage of  different wavelength ranges. The analysis was also based on an  artificial catalogue of 4970 sources built with the input parameters of the CIGALE code.  Our main results can be summarized as follows:

   \begin{enumerate}
           \item Instantaneous SFR  are found  independent of SFH assumptions with systematic differences lower than 10$\%$. The instantaneous SFR  measured by fitting the whole SED with  $\tau$-models are found fully consistent with $SFR_{\rm IR,FUV}$ measured with empirical recipes,  and $15\%$ higher  on average than $SFR_{\rm IR,FUV}$  for the two-populations models,  with a dependence on the age of the younger stellar population. The SFR  averaged over 100 Myr  and instantaneous SFR  agree well for  $\tau$-models but not for the two-populations models.   
    \item The stellar masses depend on the assumed SFH and are systematically lower  by a factor of  1.5-2  for the decl.-$\tau$ model as compared to all the other models considered. It is  caused    by  an obvious  underestimate of the age of the stellar population for the free-age decl.$\tau$ model.   Fixed-age rising-$\tau$ and free-age two-populations models are fully consistent with $M_{\rm star}$  values  lower on average by a factor of  1.3 than those obtained with  the  fixed-age two-populations model. 
      
           \item  The  IB filters sample the UV rest-frame of our galaxies, which is almost featureless. As a consequence these data are found to play a minor role, if any, in the determination of SFR, $M_{\rm star}$ as well as of stellar ages.   
       \item  Whether or not we include   NIR and IR  data modify  parameter estimations substantially .
With IR data, SFR are   measured with a dispersion of 50$\%$. Without IR data,  the  intrinsic dispersion reaches  a factor of two  and  the   range of estimated values is reduced when compared to the true values. The difference between estimations with and without  IR is tightly correlated to the uncertainty of dust attenuation measurements.  Excluding NIR data  lowers   $M_{\rm star}$  estimates  by 15$\%$ with an increase of the intrinsic dispersion from 60$\%$ with NIR data to a factor $\sim 2$ without them.  
    \item Stellar age estimates are analysed with our mock catalogue.  Systematic shifts are found: the shortest ages are overestimated and the largest ones underestimated with  a  direct impact on  $M_{\rm star}$ derivations, explaining the moderate systematic shift  and   dispersion  found between the estimated and true values of $M_{\rm star}$. 
   The impact of stellar age uncertainties on  SFR measurements is much lower,   SFR  being   far more sensitive to dust attenuation. 
     \item   These  results have some impact on  the  SFR-$M_{\rm star}$relation.  The assumption of  different SFH modifies the SFR-$M_{\rm star}$ scatter plot. When two  stellar populations are introduced, whether or not we fix the age of the oldest population has only a modest impact. The decl.-$\tau$ model leads to larger SSFR, especially for low-mass galaxies. The assumption of  a rising SFH with a fixed age  implies a well-defined SFR-$M_{\rm star}$   which does not evolve much with z.  A smaller range of SFR  is found   without IR data  as well as a  flatter  variation of the SSFR   than when IR data are introduced. 

        \end{enumerate}

\begin{acknowledgements}
This work is  supported by the French National Agency for Research (ANR-09-BLAN-0224) and  CNES. The GOODS-$Herschel$ data were accessed through the HeDaM database (http://hedam.lam.fr) operated by CeSAM and hosted by the Laboratoire d'Astrophysique de Marseille. The authors thank D. Elbaz, E. Daddi, and  M. Bethermin for very fruitful discussions, and C. Maraston and J. Pforr for their help with stellar models. The anonymous referee's suggestions have greatly helped the authors to improve the paper. 
      \end{acknowledgements}

\end{document}